\documentclass[12pt]{iopart}
\usepackage{iopams}  
\expandafter\let\csname equation*\endcsname=\relax
\expandafter\let\csname endequation*\endcsname=\relaxD
\usepackage{epsf,graphicx,graphics}
\usepackage{amsmath}
\usepackage{latexsym,amssymb}
\usepackage{setspace,cite}
\usepackage{a4}
\usepackage{slashed}
\usepackage{float}
\usepackage[left=3cm,right=3cm,top=3cm,bottom=4cm]{geometry}

\usepackage{panmaths}

\begin{document}

\title[Existence of solutions to regular topological dyonic adS EYM theories]{On the existence of topological dyons and dyonic black holes in anti-de Sitter Einstein-Yang-Mills theories with compact semisimple gauge groups}

\author{J. Erik Baxter}

\address{Dept. of Engineering and Maths,
	Sheffield Hallam University,
	Howard Street,
	Sheffield, 
	South Yorkshire, UK S1 1WB}
\ead{e.baxter@shu.ac.uk}

\begin{abstract}
	Here we study the global existence of `hairy' dyonic black hole and dyon solutions to four dimensional, anti-de Sitter Einstein-Yang-Mills theories for a general simply-connected and semisimple gauge group $G$, for so-called topologically symmetric systems, concentrating here on the \emph{regular} case. We generalise here cases in the literature which considered purely magnetic spherically symmetric solutions for a general gauge group, and topological dyonic solutions for $\mk{su}(N)$. We are able to establish the global existence of non-trivial solutions to all such systems, both near existing embedded solutions and as $|\Lambda|\rar\infty$. In particular, we can identify non-trivial solutions where the gauge field functions have no zeroes, which in the $\sun$ case proved important to stability. We believe that these are the most general analytically proven solutions in 4D anti-de Sitter Einstein-Yang-Mills systems to date.
\end{abstract}


(Accepted for publication in \emph{J. Math. Phys.} \textbf{59}:052502 (2018); doi: 10.1063/1.5000349)

\maketitle

\section{Introduction}

The study of `hairy back holes' -- black hole models with a Lagrangian which includes other matter field terms -- is now abundant in the literature, starting with when Bizon \cite{bizon_colored_1990} and Bartnik and McKinnon \cite{bartnik_particle-like_1988} published their results for non-Abelian black holes in asymptotically flat Einstein-Yang-Mills (EYM) theory, characterised by a single $\essu$-invariant gauge field $\omega$. The significance of this is that purely gravitational solutions are entirely characterised by their mass $m$, their charge $e$ and their angular momentum $a$, as proven by the uniqueness theorems of Israel, Penrose and Carter \cite{israel_event_1967,israel_event_1968,israel_300_1987}; whereas `hairy' black holes require extra degrees of freedom to entirely characterise them. For $\sun$, EYM solutions have been widely explored \cite{breitenlohner_static_1994,winstanley_existence_1999}, and indeed a wide variety of EYM solutions of various types are now known, including results for non-spherically symmetric spacetimes \cite{vanzo_black_1997, mann_topological_1997,cai_black_1996}, higher dimensions \cite{brihaye_asymptotically_2011,forgacs_particle-like_2003}, other matter fields \cite{brihaye_asymptotically_2011,yaffe_static_1989,winstanley_dressing_2005}, \emph{etc}.

The case of asymptotically anti-de Sitter (adS) space is analytically interesting for a few reasons. In asymptotically flat space, the solutions are sparse, requiring a specific discrete set of boundary values at both extremes of the spacetime, whereas adS cases possess solutions in continuous ranges of the parameter space (See e.g. \cite{winstanley_existence_1999,baxter_existence_2008,baxter_topological_2016}). This is related to stability: in adS, perturbing a solution finds a nearby solution, hence these may be stable, whereas the contrary is true for flat space. In addition, and also due to the relaxed boundary requirements, adS space allows for `dyonic solutions' possessing a non-trivial electric sector. This is in contrast to the asymptotically flat ``no magnetic charge'' case \cite{oliynyk_local_2002,kunzle_all_2002,brodbeck_self-gravitating_1994} where the electric sector must be trivial for asymptotic regularity. Since globally regular dyonic solutions are only generally supported in adS space due to the closed geometry, such solutions are less common in the literature; but monopole and dyon solutions have been found \cite{bjoraker_stable_2000}, and notably for us in the dyonic $\essu$ \cite{nolan_existence_2012,nolan_stability_2016} for which stability is proven \cite{nolan_stability_2016}, and the dyonic $\sun$ \cite{baxter_existence_2016} case. Moreover, these solutions are `nodeless' in the sense that the magnetic gauge functions possess no zeroes, which has been a necessary result for stability in previous cases \cite{lavrelashvili_remark_1995,winstanley_existence_1999,baxter_stability_2015,nolan_stability_2016}.

In this research we also consider relaxing spherical symmetry to allow for the `topological' case, investigated for $\essu$ by van der Bij \& Radu \cite{van_der_bij_new_2002}. Here we foliate spacetime by general surfaces of constant Gaussian curvature. Analytical and numerical results exist for increasingly general topological systems already \cite{baxter_existence_2015,baxter_topological_2016}. Recently published work on topological dyonic solutions in $\sun$ and for purely magnetic solutions with a general semisimple gauge group suggests there is a gap in the field that we intend to address. The author believes that this model represents the most general 4D adS EYM model which has been analytically investigated to date. The possible applications of this work mainly pertain to the adS/CFT (Conformal Field Theory) correspondence due to Maldacena \cite{maldacena_large_1998,witten_anti_1998}; for instance, results from the study of planar dyonic black holes have already been used in the study of holographic superconductors \cite{cai_introduction_2015, hendi_holographical_2016}, and higher-dimensional black holes have been connected analytically to the study of superfluids \cite{sachdev_strange_2013}. There is also the question of continuing to test Bizon's ``No-hair'' conjecture \cite{bizon_colored_1990} and possible investigation of the Black Hole Information Paradox \cite{hawking_soft_2016}; this is a subject to which we will return in Section \ref{sec:conc}.

In this paper, we shall prove the existence of black hole and soliton solutions to 4D static adS EYM field equations for general compact, simply connected, and semi-simple Lie gauge groups. We shall do this using the following general method. We shall construct our topological ans\"{a}tze before deriving the field equations themselves, reducing them down to the `regular' case \cite{oliynyk_local_2002,baxter_existence_2016}. We shall identify various embedded solutions whose existence has been proven elsewhere -- these are crucial to our proof. We shall investigate the necessary boundary conditions such that these solutions are regular at the boundaries $r=0$, $r=r_h$ and as $r\rar\infty$. We shall prove that global solutions exist to the field equations in some neighbourhood of the boundaries, and that these solutions are analytic in their boundary values. We shall demonstrate the solutions may be integrated from one boundary to the other regularly, and that when in the asymptotic regime, the solutions remain regular. Finally, we use all of this to prove that non-trivial global solutions to this system may be found \emph{a)} in some neighbourhood of a number of different existing trivial solutions, and \emph{b)} as the absolute value of cosmological constant $|\Lambda|\rar\infty$.

\section{Ans\"{a}tze for Einstein-Yang-Mills models with non-spherical symmetry}\label{sec:ansatze}

Einstein-Yang-Mills models with spherical symmetry are a relatively well-covered subject in the literature. The asymptotically flat case for the specific case of $\mk{su}(N)$ was considered in \cite{kunzle_analysis_1994} and extended to general gauge groups in \cite{brodbeck_self-gravitating_1994,oliynyk_local_2002,kunzle_all_2002}. There, it was discovered that the requirement of asymptotic regularity of the field variables was very restrictive, and necessitated that the electric sector be trivial -- i.e. these solutions are `purely magnetic', referred to as `zero magnetic charge models' \cite{brodbeck_self-gravitating_1994,brodbeck_generalized_1993}. These purely magnetic solutions have been extended to asymptotically adS space for a general gauge group in the spherically symmetric case \cite{baxter_general_2016}, and to $\mk{su}(N)$ in the case of `topological symmetry' \cite{baxter_existence_2015, baxter_existence_2016, baxter_topological_2016} (which we shall outline). However since there are no similar restrictions on the electric gauge field for adS solutions, we will extend the model to cover the cases of dyonic solutions with topological symmetry.

\subsection{Topological symmetry}

We begin by introducing the elements of the theory. Let $G$ be a compact, simply-connected, semisimple Lie group with Lie algebra $\mk{g}$. Then to demand a symmetry on the theory, we consider principal $\mc{S}$-valued automorphisms (where $\mc{S}$ is the Lie group representing the symmetry) on principal $G$-bundles $P$ over our 4D spacetime manifold $M$ with metric $g$, such that the automorphisms project onto isometry actions in $M$ whose orbits are diffeomorphic to 2-surfaces with topology $\Sigma^2$ which foliate the manifold, and where $\mbox{Aut}(\Sigma^2)=\mc{S}$.

Spherical symmetry considers the case where $\mc{S}=SU(2)$ and $\Sigma^2\cong S^2$. However we may extend the spherically symmetric $SU(N)$ system to the topological case \cite{van_der_bij_new_2002}. This generalises to three separate cases, each of which foliate spacetime by 2D surfaces of constant Gaussian curvature $K$ parametrised by `generalised angles` $(\theta,\phi)$, and which are parametrised by the sign of the curvature $k\equiv\mbox{sign}(K)$: 
\begin{enumerate}
	\item $k=1$ corresponds to the usual spherical symmetry, with Lie group $\mc{S}_1=SU(2)$ and topology $\Sigma_1^2\cong S^2$;
	\item $k=0$ corresponds to planar symmetry, with Lie group $\mc{S}_0=E(2)$ and topology $\Sigma_0^2\cong \R^2$;
	\item $k=-1$ corresponds to hyperbolic symmetry, with Lie group $\mc{S}_{-1}=SU(1,1)$ and topology $\Sigma_{-1}^2\cong H^2$;
\end{enumerate}
where we use $\Sigma_k^2$ to stand for the topology of the foliation and $\mc{S}_k$ to stand for the symmetry group in question. Each of these values of $k$ are also associated with a function $f_k(\theta)$, which endows the connection and metric tensor with the correct topology:
\begin{equation}
	f_k(\theta)=\Bigg\{
	\begin{array}{lcl}
		\sin\theta  & \mbox{for} & k=1,\\
		\theta      & \mbox{for} & k=0,\\
		\sinh\theta & \mbox{for} & k=-1.\\
	\end{array}
\end{equation}

\subsection{Metric ansatz}

As in \cite{brodbeck_self-gravitating_1994}, the induced action of $\mc{S}_k$ on $M$ is isometric, and hence we may write the metric as $g=\tilde{g}+r^2\hat{g}_k$, with $\tilde{g}$ the metric on the submanifold parametrised by $(t,r)$ and $\hat{g}_k$ the metric on the `angular' part parametrised by $(\theta,\phi)$. The same arguments there apply here, and so we conclude that it is possible to find local Schwarzschild-like topologically symmetric co-ordinates $(t,r,\theta,\phi)$ such that the metric can be written
\begin{equation}\label{metric}
	ds^2=-\mu S^2dt^2+\mu^{-1}dr^2+r^2\left(d\theta^2+f^2_k(\theta) d\phi^2\right),
\end{equation}
for $S(r)$ the lapse function and $\mu(r)$ the usual mass fraction, defined as
\begin{equation}
	\mu=k-\frac{2m}{r}+\frac{r^2}{\ell^2}
\end{equation}
where $m(r)$ is the usual mass function and $\ell$ is the adS radius of curvature $\ell=\sqrt{\frac{-3}{\Lambda}}$ (where the  cosmological constant $\Lambda<0$). Note that we are dealing solely with the static case here, implying the existence of a timelike Killing vector to which $t$ is adapted, hence all functions are of $r$ alone.

\subsection{Connection ansatz}

Now we consider the possible $G$-invariant connections on our bundle $P$ in order to derive a form for the gauge potential. The subject of possible classes of connections over principal bundles has been covered by Wang \cite{wang_invariant_1958, kobayashi_foundations_1963}, whose work in turn has been adapted to spherically symmetric connections by K\"{u}nzle \cite{kunzle_analysis_1994}. There is no distinguished action of $\mc{S}_k$ on $M$ so we must examine all conjugacy classes of such connections, which are in one-to-one correspondence with (and hence are characterised by) integral elements $W_0$ of the closed fundamental Weyl chamber $\overline{W(\Sigma)}$ of the roots of $\mk{g}$ with respect to some Cartan subalgebra $\mk{h}$ and a base $\Sigma$ \cite{bartnik_structure_1997,bartnik_spherically_1989,brodbeck_generalized_1993}. 

We let $\mk{g}_0$ be the Lie algebra of $G$, the structure group of the bundle $E$. In that case, $\mk{g}$ is equal to its complexification $(\mk{g}_0)_\C$. Also we take $\{\tau_i\}$ to be a standard basis for the Pauli matrices.
Then Wang's results \cite{wang_invariant_1958} tell us that we can write
\begin{equation}\label{WangEq1}
	W_0=2iw(\tau_3),
\end{equation}
where $w$ is the homomorphism from the isotropy group $\mc{I}_{x_0}$ of the $SU(2)$-action on $M$ at the point $x_0\in M$, determined by
\begin{equation}\label{WangEq2}
	y\ccdot\pi_0=\pi_0\ccdot w(y)\,\,\forall y\in\mc{I}_{x_0}\mbox{ if }\pi_0\in\pi^{-1}(x_0),
\end{equation}
where $\pi^{-1}(x_0)$ is the fibre above $x_0$ and the central dot notation denotes the adjoint action. Equations \eqref{WangEq1} and \eqref{WangEq2} are then known as the `Wang equations' for the system, and allow us to determine the entire gauge potential.

Given the product structure of the manifold, we write the corresponding decomposition of the gauge potential as $\mc{A}=\tilde{\mc{A}}+\hat{\mc{A}}$. In the spherical case \cite{brodbeck_self-gravitating_1994}, where $k=+1$, a gauge may always be found in which the magnetic part of the potential, $\hat{\mc{A}}$, can be written
\begin{equation}\label{Atil1}
	\hat{\mc{A}}=W_1d\theta+\left(W_2\sin\theta+W_3\cos\theta\right)d\phi.
\end{equation}
This may be derived by finding the Maurer-Cartan form for an appropriate section $\chi$ in the bundle, where for our purposes we choose
\begin{equation}\label{secsig}
	\chi=\exp(\phi\tau_3)\exp(\theta\tau_1).
\end{equation}
The result \eqref{Atil1} conveniently matches the potential we derived in \cite{baxter_existence_2015} for the purely magnetic case. We now wish to generalise this result to topological solutions.

Firstly, we notice that being fairly general in scope, Proposition 3 in \cite{brodbeck_self-gravitating_1994} carries over, so that in this case a gauge may again always be found such that the potential takes the form
\begin{equation}
	\hat{\mc{A}}=W\circ\Theta
\end{equation}
where $\Theta$ is the Maurer-Cartan form for a section $\chi$, i.e. $\Theta=\chi^{-1}d\chi$, and $W$ is the homomorphism on $G$ which is induced by $w$. We will need to apply the commutation relations for the topology in question, which in our chosen basis $\{\tau_i\}$ are
\begin{equation}\label{TopComm}
	[\tau_2,\tau_3]=\tau_1,\qquad[\tau_3,\tau_1]=\tau_2,\qquad[\tau_1,\tau_2]=k\tau_3.
\end{equation}
We begin by calculating $\Theta$:
\begin{equation}\label{Thetak01}
	\Theta=\chi^{-1} d\chi=\tau_1 d\theta+\exp(-\theta\tau_1)\tau_3\exp(\theta\tau_1)d\phi.
\end{equation}
The second term is clearly just a finite rotation of $\tau_3$ about $\tau_1$ by angle $\theta$. To deal with this term we invoke the Hadamard Lemma \cite{nestruev_smooth_2003}, giving
\begin{equation}\label{HadaTau}
	\exp(-\theta\tau_1)\tau_3\exp(\theta\tau_1)=\tau_3-\theta[\tau_1,\tau_3]+\frac{\theta^2}{2!}[\tau_1,[\tau_1,\tau_3]]-\frac{\theta^3}{3!}[\tau_1,[\tau_1,[\tau_1,\tau_3]]]+...
\end{equation}
We start with the $k=-1$ case. Using the commutators \eqref{TopComm} in \eqref{HadaTau}, we find
\begin{equation}
	\begin{split}
		\exp(-\theta\tau_1)\tau_3\exp(\theta\tau_1)
		&=\tau_3\cosh\theta+\tau_2\sinh\theta.
	\end{split}
\end{equation}
This then gives
\begin{equation}
	\Theta=\tau_1d\theta+\left(\tau_2\sinh\theta+\tau_3\cosh\theta\right)d\phi,
\end{equation}
which in turn implies that
\begin{equation}\label{Atil-1}
	\hat{\mc{A}}=W\circ\Theta=W_1d\theta+\left(W_2\sinh\theta+W_3\cosh\theta\right)d\phi,
\end{equation}
where $W_i\equiv W(\tau_i)$.

For $k=0$, beginning with \eqref{HadaTau} we find
\begin{equation}
	\begin{split}
		\exp(-\theta\tau_1)\tau_3\exp(\theta\tau_1)
		&=\tau_3+\tau_2\theta.
	\end{split}
\end{equation}
where again we have used \eqref{TopComm}. Thus,
\begin{equation}\label{Atil0}
	\hat{\mc{A}}=W_1d\theta+\left(W_2\theta+W_3\right)d\phi.
\end{equation}
To summarise, we may compile expressions (\ref{Atil1}, \ref{Atil-1}, \ref{Atil0}) into a single expression covering all three cases:
\begin{equation}
	\hat{\mc{A}}=W_1d\theta+\left(W_2f_k(\theta)+W_3\frac{df_k}{d\theta}\right)d\phi.
\end{equation}
Note that this is the same form as we derived in \cite{baxter_existence_2015} for purely magnetic topological $\mk{su}(N)$ solutions, with
\begin{equation}
	W_1=\frac{1}{2}(C-C^\dagger),\qquad W_2=\frac{-i}{2}(C+C^\dagger),\qquad W_3=\frac{i}{2}kD
\end{equation}
where $C, D\in\mk{su}(N)$ in the adjoint representation, $C$ is real and skew-hermitian and $D$ is real and diagonal.
We further note that the same arguments carry over for the (here non-trivial) electric part of the gauge potential $\tilde{A}$, and to simplify the equations we must deal with, we use a temporal gauge as in \cite{brodbeck_self-gravitating_1994}, we may again write 
\begin{equation}
	\tilde{A}=Adt
\end{equation}
for $A\in\mk{g}$. 
%
%
%
%
%
%
%
Hence, the gauge potential we shall use can be given as
\begin{equation}\label{conn}
	\mc{A}=A(r)dt+W_1(r)d\theta+\left(W_2(r)f_k(\theta)+W_3\frac{df_k}{d\theta}\right)d\phi,
\end{equation}
where we have $W_3=-\frac{i}{2}W_0$ as the constant isotropy generator, the Wang equations become
\begin{equation}\label{WE1}
	[W_3,W_1]=W_2,\qquad [W_2,W_3]=W_1,
\end{equation}
and finally, $[A,W_3]=0$ so that $A\in\mk{h}$. 

\subsection{Asymptotic regularity requirements for $\Lambda<0$}\label{sec:asymregreq}

We here take a moment to discuss the asymptotic regularity of the solutions which highlights a major difference between the cases for $\Lambda=0$ and $\Lambda<0$ -- that is, the fact that the solution space for asymptotically adS models is much larger and richer than for asymptotically flat models, due to the comparatively relaxed requirements at infinity for the former case.

In the case of $\Lambda=0$, we may \cite{oliynyk_local_2002} reduce our attention to so-called `regular' models as described in \cite{bartnik_structure_1997,brodbeck_generalized_1993}, which are those defined by conjugacy classes of bundle automorphisms that drop off sufficiently quickly at infinity (and zero, for solitons). These are described as the `zero magnetic charge models' in \cite{brodbeck_self-gravitating_1994}. Essentially this mandates that at infinity (and at the origin in the case of solitons), the magnetic gauge field functions take specific values dictated by the chosen group $G$, and also that the electric gauge field vanishes identically. This is due to the following requirements for $\Lambda=0$: let
\begin{equation}
	A^{(i)}\equiv\lim_{r\rar i}A(r),\qquad W_{j}^{(i)}\equiv\lim_{r\rar i}W_{j}(r),
\end{equation}
for $i\in\{0,\infty\}$, $j\in\{1,2\}$. Then we must have (for the topological case in hand)
\begin{equation}
	[A^{(i)},W_{j}^{(i)}]=0,\qquad [W^{(i)}_1,W^{(i)}_2]=kW_3,
\end{equation}

so that in these limits, there must be a homomorphism of $\essu$ into $\mk{g}$. This is why in asymptotically flat space, such models are sparse in the solution space. 

However, for the case of $\Lambda<0$, this is no longer true. This analysis is in Section \ref{sec:asym}, but to summarise: we transform $r$ to a variable $\tau$ that is good asymptotically, and consider the phase plane of the system essentially composed of $\left(A,W_j,\dfrac{dA}{d\tau},\dfrac{dW_j}{d\tau}\right)$. Looking for the fixed (or `critical') points of this system, where the $\tau$ derivative of each of the variables is zero and hence where phase plane trajectories end at finite values, we find a finite number of critical points, suggesting that regular solutions must approach these points asymptotically. However, the asymptotic structure of adS space is such that we may compactify the domain of integration from infinite to finite range, meaning that a solution in the variable phase space will in general not reach the end of its trajectory. Therefore the values of the critical points of the asymptotic system and the values that the field variables \emph{attain} asymptotically are not in general the same -- this is contrary to what we find in the case of $\Lambda=0$. Thus we might say that if we let $A^{(*)}$, $W_{j}^{(*)}$ be the values of $A$ and $W_{j}$ at these critical points, we still have the altered constraints
\begin{equation}\label{asymcon}
	\begin{split}
		[A^{(0)},W_\pm^{(0)}]=[A^{(*)},W_{\pm}^{(*)}]=0,\qquad [W^{(0)}_1,W^{(0)}_2]=[W^{(*)}_1,W^{(*)}_2]=kW_3,
	\end{split}
\end{equation}
but that in general
\begin{equation}
	A^{(\infty)}\neq A^{(*)},\qquad W_{j}^{(\infty)}\neq W_{j}^{(*)}
\end{equation}
for $j\in\{1,2\}$. Hence, in adS space, solutions are far more plentiful, and it is this that enables us to prove the existence of global solutions in these cases. Thus as in the purely magnetic case, the definition of a `regular model' must be extended in light of the above.

\section{Deriving the field equations in the regular case}\label{sec:FEs}

We have reduced the possible conjugacy classes of our bundle automorphisms to meet the requirement of boundary regularity, but we still have one possible action of $SU(2)$ on the bundle $E$ for each element in $\overline{W(\Sigma)}\cap I$, the intersection of the closed fundamental Weyl chamber and the integral lattice defined by $I\equiv\ker(\exp|_\mk{h})$. This is still a countably infinite quantity of possible actions. For regular models however, the constraints \eqref{asymcon} must still be obeyed by $W_0$ (and hence $W_3$) and therefore $W_0$ must still be an $A_1$-\emph{vector}, i.e. the defining vector of an $\mk{sl}(2)$-subalgebra. This set of vectors is finite and has been tabulated for the simple Lie groups by Dynkin and Mal'cev \cite{malcev_commutative_1945,dynkin_semisimple_1952}. These `characteristics' are in one-to-one correspondence with strings of integers from the set $\{0,1,2\}$, which define the values of the simple roots on $W_0$ chosen so that it lies in $\overline{W(\Sigma)}$, and so provides a classification of all possible models, including these topological ones, which obey appropriate regularity requirements at one or both boundaries for any semisimple compact gauge group.

\subsection{Field Equations}

In this Section we will reduce the field equations for $\Lambda<0$ to the regular case, and show that as in the purely magnetic and asymptotically flat cases, these models coincide with those for the principal action, for any semisimple gauge group. We will show that such models can be entirely characterised by 2 real functions, $m(r)$ and $S(r)$, and $2\mc{L}$ real functions of $r$ representing the gauge fields.

The EYM field equations are well-known:
\labeq{\begin{split}
	& 2T_{\mu\nu}=G_{\mu\nu}+\Lambda g_{\mu\nu},\\& 0=\nabla_\lambda F^{\,\,\lambda}_{\mu}+[A_\lambda,F^{\,\,\lambda}_\mu],\\
\end{split}}{GenFEs}
where $g_{\mu\nu}$ is the metric tensor defined using \eqref{metric}, $G_{\mu\nu}$ is the Einstein tensor, $F^{\,\,\lambda}_\mu$ is the mixed anti-symmetric field strength tensor defined with
	\labeq{F_{\mu\nu}=\partial_\mu A_\nu-\partial_\nu A_\mu + [A_\mu,A_\nu],}{Fmunu}
$A_\mu$ is the Yang-Mills one-form connection \eqref{conn} given by $\mc{A}=A_\mu dx^\mu$, and the energy-momentum tensor $T_{\mu\nu}$ is given by
	\labeq{T_{\mu\nu}\equiv\mbox{Tr}\left[F_{\mu\lambda}F^{\,\,\lambda}_\nu-\frac{1}{4}g_{\mu\nu}F_{\lambda\sigma}F^{\lambda\sigma}\right].}{Tmunu}
We note that Tr is the Lie algebra trace, we have used the Einstein summation convention where summation occurs over repeated indices, and we have rescaled all units so that $4\pi G=c=q=1$ (for the gauge coupling constant $q$).

A more convenient basis to use here for the Wang equations \eqref{WE1} in place of the generators $W_1$ and $W_2$ is
\begin{equation}\label{WD}
	W_\pm=\mp W_1-iW_2,
\end{equation}
in which case equations \eqref{WE1} become
\begin{equation}\label{WE2}
	[W_0,W_\pm]=\pm 2W_\pm.
\end{equation}
Then $W_\pm(r)$ are $\mk{g}$-valued functions, $W_0$ is a constant vector in a fundamental Weyl chamber of $\mk{h}$, and $\{W_0,W_\pm\}$ is a standard $\essu$ triple in the limit $r=0$ and at the critical points of the system (See Section \ref{sec:asym}). Also, $\mk{h}$ is the Cartan subalgebra of the complexified form of the Lie algebra, i.e. $\mk{h}=\mk{h}_0+i\mk{h}_0$, for $\mk{h}_0$ the real Cartan subalgebra of $\mk{g}_0$, which in turn is the real compactified form of $\mk{g}$. We introduce a complex conjugation operator $c:\mk{g}\rar\mk{g}$ such that
\begin{equation}
	c(X+iY)=X-iY,\,\,\forall X,Y\in\mk{g}_0,
\end{equation}
which implies 
\begin{equation}
	W_-=-c(W_+).
\end{equation}
This is consistent with us having written the field equations such that $c(\mc{A})=\mc{A}$. 

Using \eqref{metric}, \eqref{conn} and \eqref{GenFEs}, we may derive the field equations. Defining the following quantities:
\begin{equation}\label{QuantDefsComm}
	\begin{array}{ll}
		\eta=-\frac{1}{2}(A^\prime,A^\prime),\qquad & \hat{F}=\frac{i}{2}\left(kW_0-[W_+,W_-]\right),\\ & \\
		G=\frac{1}{2}(W_+,W_-),	& \zeta=-\frac{1}{2}\left([A,W_+],[A,W_-]\right), \\ & \\
		\mc{F}=-i[\hat{F},W_+], & \mc{Z}=[W_+,[A,W_-]],\\ & \\
		P=-\frac{1}{2}(\hat{F},\hat{F}), & \\
	\end{array}
\end{equation}
we find the Einstein equations to be
\begin{subequations}\label{EEs}
	\begin{align}
	m^\prime&=\frac{r^2\eta}{S^2}+\frac{\zeta}{\mu S^2}+\mu G+\frac{P}{r^2},\label{EEm}\\
	\frac{S^\prime}{S}&=\frac{2G}{r}+\frac{2\zeta}{\mu^2 S^2 r},\label{EES}
	\end{align}
\end{subequations}
and the (non-zero, independent) Yang-Mills equations can be written
\begin{subequations}\label{CommEqns}
	\begin{align}
	0&=[W_+,W^\prime_-]-[W^\prime_+,W_-],\label{YMWWComm}\\
	0&=A^{\prime\prime}+\left(\frac{2}{r}-\frac{S^\prime}{S}\right)A^\prime-\frac{\mc{Z}}{\mu r^2},\label{YMAComm}\\
	0&=r^2\mu W^{\prime\prime}_++r^2\left(\mu^\prime+\mu\frac{S^\prime}{S}\right)W^\prime_+-\frac{r^2}{\mu S^2}[A,[A,W_+]]+\mc{F}.\label{YMWComm}
	\end{align}
\end{subequations}
Above we have used an invariant inner product $(\,\, , \,)$ on $\mk{g}$, which we will define properly below. This arises from the Lie algebra trace, and is determined up to a factor on each simple component of a semisimple $\mk{g}$. This inner product induces a norm $|\,\,|$ on (the Euclidean) $\mk{h}$ and therefore also on its dual, and the factors are chosen so that $(\,\, , \,)$ is a positive multiple of the Killing form $\langle\,\, ,\,\,\rangle$ on each simple component. Also, we note that since $c(\hat{F})=\hat{F}$, $c(A)=A$ and $\HIP{X}{Y}\equiv-(c(X),Y)$ is a Hermitian inner product on $\mk{g}$, then the quantities $\eta,\zeta,G,P\geq 0$. Finally then, we may calculate the energy density $e$, which is an quantity we will need when we simplify our solutions by considering only those with the correct asymptotic behaviour, noting that as expected it is non-negative:
\begin{equation}\label{EnergyDensity}
	e=r^{-2}\left(\frac{r^2\eta}{S^2}+\frac{\zeta}{\mu S^2}+\mu G+\frac{P}{r^2}\right).
\end{equation}
%

Now we reduce the field equations down to the case of a regular action by choosing an explicit Chevally-Weyl basis for $\mk{g}$. Let $R$ be the set of roots on $\mk{h}^*$ and $\Sigma=\{\alpha_1,...,\alpha_\mc{L}\}$ be a basis for $R$ (where $\mc{L}$ is the rank of $\mk{g}$). To define the inner product $(\,\, , \,\,)$, we make the definitions
\begin{equation}\label{innerproddef}
	(\bv{t}_\alpha,X)\equiv\alpha(X)\quad\forall X\in\mk{h},\qquad 	\bv{h}_\alpha\equiv\frac{2\bv{t}_\alpha}{|\alpha|^2}.
\end{equation}
We let $\{\bv{h}_i\equiv\bv{h}_{\alpha_i},\bv{e}_\alpha,\bv{e}_{-\alpha}\,|\,i=1,...,\mc{L};\,\alpha\in R\}$ be a basis for $\mk{g}$. This basis induces the natural decomposition
\begin{equation}
	\mk{g}=\mk{h}\oplus\bigoplus_{\alpha\in R^+}\mk{g}_\alpha\oplus\mk{g}_{-\alpha}
\end{equation}
where $R^+$ are the set of positive roots expressed in the basis $\Sigma$. For this decomposition, we take the conventions
\begin{equation}\label{basedef}
	[\bv{e}_{\alpha},\bv{e}_{-\alpha}]=\bv{h}_{\alpha},\qquad [\bv{e}_{-\alpha},\bv{e}_{-\beta}]=-[\bv{e}_{\alpha},\bv{e}_{\beta}],\qquad(\bv{e}_{\alpha},\bv{e}_{\beta})=\frac{2\bsym{\delta}^\alpha_{-\beta}}{|\alpha|^2}.
\end{equation}
for $\bsym{\delta}^\alpha_\beta$ the Kronecker symbol. We define an $\mk{sl}(2)$-subalgebra $\mbox{span}\{\bv{e}_{0},\bv{e}_{\pm}\}$ of $\mk{g}$ using the appropriate commutators, i.e.
\begin{equation}
	[\bv{e}_{0},\bv{e}_{\pm}]=\pm2\bv{e}_{\pm},\qquad[\bv{e}_{+},\bv{e}_{-}]=\bv{e}_{0};
\end{equation}
and if we notice that
\begin{equation}
	[\bv{h}_{},\bv{e}_{\alpha}]=\alpha(\bv{h}_{})\bv{e}_{\alpha},
\end{equation}
then according to \cite{dynkin_semisimple_1952}, $\bv{e}_{0}$ can only be an $A_1$-vector if there exists an $\alpha\in R$ such that
\begin{equation}
	\alpha(\bv{e}_0)=2. 
\end{equation}

Using these facts then, we may write $W_0\in\mk{h}$ in the basis
\begin{equation}
	W_0=\slim_{i=1}^\mc{L}w_i\bv{h}_{i}\in\mk{h},
\end{equation}
so that equations \eqref{WE2} imply that
\begin{equation}\label{Wplus}
	W_+(r)=\slim_{\alpha\in \Sigma_w}\omega_\alpha(r)\bv{e}_{\alpha},\quad 	W_-(r)=\slim_{\alpha\in \Sigma_w}\varpi_\alpha(r)\bv{e}_{-\alpha}
\end{equation}
for two sets of complex functions $\omega_\alpha$ and $\varpi_\alpha$, where we have defined $\Sigma_w$, a set of roots depending on the homomorphism $w$ -- i.e. the constants $w_i$ -- as
\begin{equation}\label{slamdef}
	\Sigma_w\equiv\{\alpha\in R\,|\,\alpha(W_0)=2\}.
\end{equation}
However, given that the complex conjugation operator $c$ maps
\begin{equation}\label{cmap}
	\bv{h}_{i}\mapsto-\bv{h}_{i},\qquad\bv{e}_{\alpha}\mapsto -\bv{e}_{-\alpha},
\end{equation}
it is clear that
\begin{equation}
	\varpi_\alpha(r)=c(\omega_{\alpha}(r)),
\end{equation}
reducing the number of independent functions we have. As for the electric sector, since $[A,W_3]=0$, then $A\in\mk{h}$, and we can write $A$ in the form
\begin{equation}
	A=i\slim_{j=1}^\mc{L}a_j(r)\bv{h}_j.
\end{equation}
This is because $c(A)=A$, so that \eqref{cmap} implies the functions $a_i(r)$ are purely real. In addition, the first Yang-Mills equation \eqref{YMWWComm} can be solved using the same argument as in \cite{baxter_general_2016,oliynyk_local_2002}, implying that $\omega_j(r)\in\R\,\,\forall j$. We note that in $\Lambda=0$ purely magnetic solutions, this is only possible for the regular case \cite{oliynyk_local_2002}. Therefore, the system is determined by two real functions $m(r)$, $S(r)$, and $2\mc{L}$ real functions $a_i(r)$, $\omega_i(r)$, $\forall i\in\{1,...,\mc{L}\}$.

We further note that as in \cite{brodbeck_self-gravitating_1994}, we must also consider the form of $A$, since it still may be expressed as the direct sum of two orthogonal sectors. The element $A$ is valued in $LT$, the infinitesimal torus which is the centraliser of $G$. Therefore, relative to the decomposition $LT=\langle \Sigma_w\rangle\oplus\langle\Sigma_w\rangle^\perp$, where here $\langle\,\rangle$ indicates the span, we can write $A=A_\parallel+A_\perp$. Note that this decomposition is independent of our choice of $\mbox{Ad}\,G$-invariant inner product. We wish to examine the form of $A_\perp$. The cases of solitons and black holes must be taken separately.

For solitons, the argument in \cite{brodbeck_self-gravitating_1994} applies identically here, and shows that $A_\perp$ may be gauged away. Examining the perpendicular component of \eqref{YMAComm}, noting that again $[W_+,[A,W_-]]\in\langle \Sigma_w \rangle$, then we find $r^2 S^{-1} A_\perp'=c$, for a constant arbitrary vector $c\in\langle\Sigma_w\rangle^\perp$. For the energy density \eqref{EnergyDensity} to be finite at the origin, then we must have $c=\ul{0}$ so that $A_\perp$ is a constant. Finally, the gauge transform $A\rar (\mbox{ad}(g))^{-1}A+g^{-1}dg$ with $g=\exp(-A_\perp t)$ gauges $A_\perp$ to zero but leaves the rest of the potential unaffected. 

For black holes, we still have $r^2 S^{-1}A_\perp'=c$, but since $r\geq r_h$, regularity does not immediately imply that $c=\ul{0}$. Hence, as in \cite{brodbeck_instability_1996}, we adopt a gauge where $\lim_{r\rar\infty}A_\perp=0$, and we obtain the integral solution
\begin{equation}\label{APerp}
	A_\perp(r)=c\int\limits_{r}^\infty\frac{S(y)}{y^2}dy.
\end{equation}
This matches what is found in \cite{brodbeck_instability_1996}. 
Noting also that $[\langle\Sigma_w\rangle^\perp,W_\pm]=0$, then \eqref{APerp} decouples from the rest of the equations and thus $A_\perp$ can be found when $S(r)$ is known.

Hence, from now on we simply work with $A$; knowing that for solitons, $A=A_\parallel$, and for black holes, once we have found a solution to the field equations, we can use knowledge of $S(r)$ in \eqref{APerp} to `separate' $A_\parallel$ from $A_\perp$. This is perhaps why the term has not been mentioned in previous treatments of $\sun$ dyonic black holes \cite{nolan_existence_2012,nolan_stability_2016,baxter_existence_2016}. We should mention that the rest of the proof in \cite{brodbeck_self-gravitating_1994}, where also $A_\parallel=0$ for regularity, fails because of the altered asymptotic requirements for $\Lambda<0$ (Section \ref{sec:asym}).

\subsection{Regular models}

It is noted in \cite{oliynyk_local_2002} that we may simplify the system a lot by considering only the $\textit{regular}$ case, where $W_0$ is a vector in the \emph{open} fundamental Weyl chamber $W(S)$ \cite{brodbeck_self-gravitating_1994}. We begin with an extension of a theorem from Brodbeck and Straumann:
\begin{thr} \cite{brodbeck_self-gravitating_1994}\label{Kthm1}
	If $W_0$ is in the open Weyl chamber $W(\Sigma)$ then the set $\Sigma_w$ is a $\Pi$-system, i.e. satisfies:
	
	\begin{enumerate}
		\item if $\alpha,\beta\in \Sigma_w$ then $\alpha-\beta\notin R$,\\
		\item $\Sigma_w$ is linearly independent;\\
		
		and is therefore the base of a root system $R_w$ which generates a Lie subalgebra $\mk{g}_w$ of $\mk{g}$ spanned by $\{\bv{\emph{h}}_{\alpha},\bv{\emph{e}}_{\alpha},\bv{\emph{e}}_{-\alpha}\,|\,\alpha\in R_w\}$. Moreover, if $\mk{h}_w\equiv\mbox{\emph{span}}\{\bv{\emph{h}}_{\alpha}\,|\,\alpha\in \Sigma_w\}$ and $\mk{h}^\perp_w\equiv\bigcap_{\alpha\in \Sigma_w}\ker\alpha$ then
		\begin{equation}
		\mk{h}=\mk{h}_w^\parallel\oplus\mk{h}^\perp_w\quad\mbox{ and }\quad W_0=W_0^\parallel + W_0^\perp\quad\mbox{ with }\quad W_0^\parallel=\slim_{\alpha\in R_w}\bv{\emph{h}}_{\alpha}.
		\end{equation}
		
		If $W_0$ is an $A_1$-vector then $W_0^\perp=0$ (though $\mk{h}_w^\perp$ need not be trivial).\\
		
		To this theorem we may add the extra point from above, only applicable to $\Lambda<0$:
		%
		\item Let $A=A_\parallel+A_\perp$. Then for solitons, we may find a gauge in which $A_\perp=0$, and for black holes, we can at least show that $A_\perp$ decouples and can be found when we know $S(r)$. We also note that $A_\parallel$ is \ul{not} required to vanish, unlike the asymptotically flat case.
	\end{enumerate}
\end{thr}
As in the purely magnetic case, we will see that this reduces the equations down to a form resembling the $\sun$ equations studied in \cite{baxter_existence_2016}. First we can consider $W_+$ to be a $\mk{g}_w$-valued function, and write
\begin{equation}\label{wBasis}
	W_{+}(r)=\slim_{j=1}^{\mc{L}_{w}}\omega_j(r)\tilde{\bv{e}}_j,\quad 	A=i\slim_{j=1}^{\mc{L}_w}a_j(r)\tilde{\bv{h}}_j,
\end{equation}
where we now take $\{\tilde{\alpha}_1,...,\tilde{\alpha}_{\mc{L}_w}\}$ as the basis for $\Sigma_w$ and define $\tilde{\bv{e}}_j\equiv\bv{e}_{\tilde{\alpha}_j}$. 
%
%
%
The functions $a_j$ are purely real. As in the dyonic $\sun$ case \cite{baxter_existence_2015}, we find it more convenient here to write the electric field in a different basis, defined by
\begin{equation}\label{Ebar}
	\begin{split}
		\overline{E}_+&\equiv-i[A,\tilde{\bv{e}}_{\alpha_j}]
		\equiv \slim_{j=1}^{\mc{L}_w}\mc{E}_j\tilde{\bv{e}}_{\alpha_j},\,\,\mbox{i.e.}\,\,\mc{E}_j=\slim_{l=1}^{\mc{L}_w}a_l(r)\tilde{\alpha}_j(\tilde{\bv{h}}_l)
	\end{split}
\end{equation}
for $\mc{L}_w$ real functions $\mc{E}_i(r)$. This conveniently sets both equations in the same basis of vectors. Also using this basis, we may define the Cartan matrix of the reduced subalgebra $\mk{g}_w$ as
\begin{equation}\label{Cartandef}
	C_{ij}\equiv\langle\tilde{\alpha}_i,\tilde{\alpha}_j\rangle,
\end{equation}
noting that by definition this is a symmetric and positive operator. In addition, we are now in a position to elucidate the structure of our inner product a little: We note that in this basis for the roots, then due to the definition of the Killing form, 
%
%
and using \eqref{innerproddef} and \eqref{Cartandef}, we can derive that
\begin{equation}\label{KillingH}
	\begin{split}
		(\tilde{\bv{h}}_i,\tilde{\bv{h}}_j)
		=\frac{2}{|\tilde{\alpha}_j|^2}C_{ij},
	\end{split}
\end{equation}
a relation that will be useful to us momentarily.

Now we rewrite the remaining Yang-Mills equations. The field equation for the electric sector \eqref{YMAComm} easily becomes
\begin{equation}
	\slim_{i=1}^{\mc{L}_w}\left[a_i^{\prime\prime}+\left(\frac{2}{r}-\frac{S^\prime}{S}\right)a^\prime_i-\frac{\omega^2_i}{\mu r}\slim_{j=1}^{\mc{L}_w}a_j\tilde{\alpha}_i(\tilde{\bv{h}}_j)\right]\tilde{\bv{h}}_i=0,
\end{equation}
and if we take the commutator of this with $\tilde{\bv{e}}_{\alpha_k}$ and sum over $k$, we get the $\mc{L}_w$ equations
\begin{equation}
	\mc{E}_i^{\prime\prime}+\left(\frac{2}{r}-\frac{S^\prime}{S}\right)\mc{E}^\prime_i-\frac{1}{\mu r}\slim_{j=1}^{\mc{L}_w}C_{ij}\mc{E}_j\omega^2_j=0.
\end{equation}
%

This leaves the magnetic equations \eqref{YMWComm}. Using the first equation in \eqref{basedef}, we can compute $\hat{F}$ to be
\begin{equation}\label{F}
	\begin{split}
		\hat{F}=&\dfrac{i}{2}\left[\slim_{i=1}^{\mc{L}_w}k\lambda_i\bv{h}_i-\left[\slim_{i=1}^{\mc{L}_w}\omega_i\tilde{\bv{e}}_i,\slim_{i=j}^{\mc{L}_w}\omega_j\tilde{\bv{e}}_j\right]\right]\\
		=&\dfrac{i}{2}\slim_{i=1}^{\mc{L}_w}\left(k\lambda_i-\omega_i^2\right)\tilde{\bv{h}}_i,
	\end{split}
\end{equation}
then using this, \eqref{innerproddef} and \eqref{KillingH}, we can calculate $P$ as
\begin{equation}\label{P}
	\begin{split}
		P=&\dfrac{1}{8}\left(\slim_{i=1}^{\mc{L}_w}(k\lambda_i-\omega_i^2)\tilde{\bv{h}}_i,\slim_{j=1}^{\mc{L}_w}(k\lambda_j-\omega_j^2)\tilde{\bv{h}}_j\right)\\
		=&\dfrac{1}{8}\slim_{i,j=1}^{\mc{L}_w}(k\lambda_i-\omega_i^2)h_{ij}(k\lambda_j-\omega_j^2)
	\end{split}
\end{equation}
where
\begin{equation}\label{hdef}
	h_{ij}\equiv\disfrac{2C_{ij}}{|\tilde{\alpha}_j|^2}.
\end{equation}
Also, 
\begin{equation}\label{AAW}
	[A,[A,W_+]]=-\left[\slim_{i=1}^{\mc{L}_w}a_i(r)\tilde{\bv{h}}_i\,\, ,\slim_{j=1}^{\mc{L}_w}\mc{E}_j\omega_j\tilde{\bv{e}}_{\alpha_j}\right]=-\slim_{i=1}^{\mc{L}_w}\mc{E}_i^2\omega_i\tilde{\bv{e}}_{\alpha_i}.
\end{equation}
Hence, the $\mc{L}_w$ magnetic gauge field equations \eqref{YMWComm} remain the same as in the purely magnetic case \cite{baxter_general_2016} with the addition of the term \eqref{AAW}, and we obtain
\begin{equation}
	\omega_j^{\prime\prime}+\left(\frac{\mu^\prime}{\mu}+\frac{S^\prime}{S}\right)\omega_j^\prime+\frac{\mc{E}_j^2\omega_j}{\mu^2 S^2}+\frac{\omega_j}{2\mu r^2}\slim_{i=1}^{\mc{L}_w}C_{ij}\left(kw_i-\omega_i^2\right)=0.
\end{equation}

Now for the Einstein equations \eqref{EEs}. The quantities $G$ and $P$ are also the same as in the purely magnetic case \cite{baxter_general_2016}, and we need only consider $\zeta$ and $\eta$:
\begin{equation}\label{QuantDefsSum}
	\begin{array}{rlrl}
		\zeta & =-\disfrac{1}{2}([A,W_+],[A,W_-]) & \eta & =-\disfrac{1}{2}(A',A')\\[5pt]
		& =\disfrac{1}{2}\slim_{i,j=1}^{\mc{L}_w}\mc{E}_i\mc{E}_{j}\omega_i\omega_j\left(\tilde{\bv{e}}_{\alpha_i},\tilde{\bv{e}}_{\alpha_{-j}}\right) & \qquad & =\disfrac{1}{2}\slim_{i,j=1}^{\mc{L}_w}a^\prime_i a^\prime_j\left(\tilde{\bv{h}}_i,\tilde{\bv{h}}_j\right)\\
		&=\slim_{i=1}^{\mc{L}_w}\disfrac{\mc{E}_i^2\omega_i^2}{|\alpha_i|^2}; & \qquad & =\disfrac{1}{2}\slim_{i,j=1}^{\mc{L}_w}a_i^{\prime}h_{ij}a_j^\prime.\\
	\end{array}
\end{equation}
We note that $\eta$ can also be written in our new basis $\overline{E}_+$ \eqref{Ebar} as
\begin{equation}\label{etaDefSum}
	\eta=\slim_{i,j=1}^{\mc{L}_w}\mc{E}_i^{\prime}\frac{(C^{-1})_{ij}}{|\alpha_j|^2}\mc{E}_j^\prime,
\end{equation}
with $(C^{-1})_{jk}$ being the matrix inverse of $C_{jk}$, which therefore is also positive and symmetric. This is due to \eqref{basedef} which implies for the regular case that in our Chevally-Weyl basis, $[\tilde{\bv{h}}_i,\tilde{\bv{e}}_{\alpha_j}]=C_{ij}\tilde{\bv{e}}_{\alpha_j}$, and thus
\begin{equation}
	\mc{E}_i=\slim_{j=1}^{\mc{L}_w}C_{ij}a_j.	
\end{equation}
Finally then, we can present our field equations for regular models. The Einstein equations are as above \eqref{EEs}, with $\zeta$ and $\eta$ given in (\ref{QuantDefsSum}, \ref{etaDefSum}), and
\begin{subequations}\label{MagQuantDefs}
	\begin{align}
	P&=\disfrac{1}{8}\slim_{i,j=1}^{\mc{L}_w}(kw_i-\omega_i^2)h_{ij}(kw_j-\omega_j^2),\label{Pdef2}\\ G&=\slim_{i=1}^{\mc{L}_w}\disfrac{\omega_i^{\prime 2}}{|\alpha_i|^2},\\
	\mc{F}_i&=\disfrac{\omega_i}{2}\slim_{j=1}^{\mc{L}_w}C_{ij}(kw_j-\omega^2_j),\\ \mc{Z}_i&=\slim_{j=1}^{\mc{L}_w}C_{ij}\mc{E}_j\omega^2_j,
	\end{align}
\end{subequations}
and $h_{ij}$ given by \eqref{hdef}. The $2{\mc{L}_w}$ Yang-Mills equations are these, where we have rewritten them using the Einstein equations \eqref{EEs}:
\begin{subequations}\label{YMEs}
	\begin{align}
	0&=\mc{E}^{\prime\prime}_i+\frac{2}{r}\left(1-G-\frac{\zeta}{\mu^2 S^2}\right)\mc{E}^\prime_i-\frac{\mc{Z}_i}{\mu r^2},\label{YMM}\\
	0&=r^2\mu\omega_i^{\prime\prime}+2\left(m-\frac{P}{r}-\frac{r^3\eta}{S^2}+\frac{r^3}{\ell^{2}}\right)\omega^\prime_i+\frac{r^2\mc{E}_i^2\omega_i}{\mu S^2}+\mc{F}_i.\label{YME}
	\end{align}
\end{subequations}
The final step is to determine the values of the constants $w_j$, which involves determining the subalgebra $\mk{g}_w$ for a given $A_1$-vector $W_0$ in the open fundamental Weyl chamber. For a semisimple group, for which the Cartan subalgebra splits into the direct sum of (orthogonal) space $\mk{h}=\bigoplus_k\mk{h}_k$, then the orthogonal decomposition given in Theorem \ref{Kthm1} splits into analogous decompositions of each of $\mk{h}_k$. Hence we only need consider the regular actions of simple Lie groups.

However, we note that the $A_1$-vector in the Cartan subalgebra $\mk{h}$ of a Lie algebra $\mk{g}$ is uniquely determined by the integers
\begin{equation}\label{nudef}
	\{\nu_1,...,\nu_\mc{L}\}\equiv\{\alpha_1(W_0),...,\alpha_\mc{L}(W_0)\}
\end{equation}
chosen from the set $\{0,1,2\}$. In \cite{dynkin_semisimple_1952}, this set is referred to as the \emph{characteristic}. From \eqref{slamdef}, it is obvious that for the principal action, 
\begin{equation}\label{nudefprin}
	\nu_j=2\quad(\forall j\in\{1,...,\mc{L}\})
\end{equation}
for $\mk{h}_w$. $A_1$-vectors satisfying this define \emph{principal $\essu$-subalgebras}, and hence \emph{principal actions} of $SU(2)$ on the bundle. Therefore may rely on the following theorem, the gist of which is that the regular action coincides with the principal action:
\begin{thr}\cite{oliynyk_local_2002}\label{Kthm2}
	\begin{enumerate}
		\item The possible regular $\essu$-subalgebras of simple Lie algebras consist of the principal subalgebras of all Lie algebras $A_\mc{L}$, $B_\mc{L}$, $C_\mc{L}$, $D_\mc{L}$, $G_2$, $F_4$, $E_6$, $E_7$ and $E_8$ and of those subalgebras  of $A_\mc{L}=\mk{sl}(\mc{L}+1)$ with even $\mc{L}$ corresponding to partitions $[\mc{L}+1-k,k]$ for any integer $k=1,...,\mc{L}/2$, or, equivalently, characteristic (22...2211...1122...22) ($2k$ `1's in the middle and `2's in all other positions);\\
		\item The Lie algebra $\mk{g}_w$ is equal to $\mk{g}$ in the principal case, and for $A_\mc{L}$ with even $\mc{L}$ equal to $A_{\mc{L}-1}$ for $k=1$ and to $A_{\mc{L}-k}\oplus A_{k-1}$ for $k=2,...,\mc{L}/2$;\\
		\item In the principal case $\mk{h}_w^\parallel=\mk{h}$. For all $\essu$-subalgebras of $A_\mc{L}$ with even $\mc{L}$ the orthogonal space $\mk{h}_w^\perp$ is one-dimensional.
	\end{enumerate}
\end{thr}
In light of the above, we now drop tildes and $w$-subscripts for clarity. Finally then we may determine an expression for the constants $w_j$, derived by using (\ref{hdef}, \ref{Pdef2}, \ref{nudef}, \ref{nudefprin}):
\begin{equation}\label{lamdef}
	w_j=2\slim_{k=1}^\mc{L}(C^{-1})_{jk}.
\end{equation}

\subsection{Trivial solutions}\label{sec:embed}

In Section \ref{sec:GloExArg}, we will argue the existence of global solutions in some neighbourhood of existing embedded (or `trivial') solutions. Therefore, we here review some known trivial solutions to the field equations \eqref{EEs} and \eqref{YMEs}.

\subsubsection{Schwarzschild anti-de Sitter (SadS)}\label{sec:SadS}

Here we notice that let a solution is found if we let $\omega_i(r)\equiv \sqrt{kw_i}$, $a_i(r)\equiv 0\,\forall r,i$ (implying also that $\mc{E}_i\equiv 0\,\forall r,i$). Note though, that this solution is only really valid for $k=1$ (so $\omega_i(r)\equiv \sqrt{w_i}$), because for $k=0$ we instead get the Reissner-N\"{o}rdstrom-adS solution (See Section \ref{sec:RNadS} below), and for $k=-1$, $\omega_i\notin\R$. Substituting into the defined quantities \eqref{QuantDefsSum}, \eqref{MagQuantDefs} we find that $\eta=\zeta=P=G=0$ and $\mc{F}_i=\mc{Z}_i=0\,\,\,\forall i$. This implies \emph{a)} that $m^\prime(r)=0$ from \eqref{EEm},  so that $M$ is a constant which we set to the ADM mass; \emph{b)} that we have $S^\prime(r)=0$ from \eqref{EES}, so that $S$ is a constant which we scale to $1$ for the asymptotic limit; and \emph{c)} the Yang-Mills equations \eqref{YMEs} are automatically satisfied. Since $\eta=P=0$, this solution carries no global charge. This can thus be identified as the embedded Schwarzschild-anti-de Sitter solution. 

\subsubsection{Reissner-N\"{o}rdstrom anti-de Sitter (RNadS)}\label{sec:RNadS}

Here we let $\omega_i(r)\equiv\mc{E}_i(r)\equiv0$. In that case, again we find $\zeta=G=0$ and therefore $S(r)$ becomes a constant, which we scale to 1. Also, $\eta=0$, and so the metric function $\mu(r)$ becomes
\begin{equation}
	\mu=k-\frac{2M}{r}+\frac{Q^2}{r^2}+\frac{r^2}{\ell^2},
\end{equation}
Again, $M$ is the ADM mass of the solution, and the magnetic charge $Q$ is defined with
\begin{equation}\label{Qdef}
	Q^2\equiv 2P=\frac{k^2}{4}\slim_{i,j=1}^\mc{L}w_ih_{ij}w_j.
\end{equation}
Thus we have the embedded Reissner-N\"{o}rdstrom anti-de Sitter solution, which we note only exists with this value of $Q^2$. Note that the hyperbolic and spherical cases carry the same charge, and the planar case necessarily has zero magnetic charge, similar to the $\Lambda=0$ case. We also note that since $P\geq0$, $Q^2\geq0$ always. 

\subsubsection{Embedded Reissner-N\"{o}rdstrom Abelian solutions}

Here we let $\omega_i\equiv 0$, and
\begin{equation}
	a_i(r)\equiv\frac{\tilde{a_i}}{r},
\end{equation}
for all $i$, where $\tilde{a}_i$ are arbitrary real constants. Then we have
\begin{equation}
	m(r) = M-\frac{1}{2r}\left(\frac{1}{2}\slim_{j=1}^\mc{L}\tilde{a}_ih_{ij}\tilde{a}_j+\frac{k^2}{4}\slim_{i,j=1}^\mc{L}w_ih_{ij}w_j\right).
\end{equation}
Identifying the bracketed terms as the electric and magnetic charges respectively, we let
\begin{equation}
	Q_E^2=\,\frac{1}{2}\slim_{j=1}^\mc{L}\tilde{a}_ih_{ij}\tilde{a}_j,\qquad Q_M^2=\,\frac{k^2}{4}\slim_{i,j=1}^\mc{L}w_ih_{ij}w_j.
\end{equation}
Then we may write the metric function $\mu(r)$ as
\begin{equation}
	\mu=k-\frac{2M}{r}+\frac{Q_E^2+Q_M^2}{r^2}+\frac{r^2}{\ell^2}.
\end{equation}
This solution is essentially an embedded $[\mk{u}(1)]^\mc{L}$ solution, which lives entirely in $\langle LT\rangle$. The arbitrariness of the constants $\tilde{a}_i$ is something that is familiar from the $\essu$ \cite{nolan_existence_2012} and $\sun$ cases \cite{baxter_existence_2016}.

\subsubsection{Embedded $\essu$ solutions}\label{sec:su2embed}

Noting that we can embed $SU(2)$ isomorphically into any semisimple gauge group $G$, then there must always exist trivial embedded $\essu$ solutions to the field equations. We may show this by a simple rescaling.

Consider the gauge group $G$, fixing the symmetry action such that $W_0$ is regular. Select any basis such that the set $\{W_0,\Omega_+,\Omega_-\}$ spans $\essu$, with $c(\Omega_+)=-\Omega_-$. We rescale the field variables as follows:
\begin{equation}\label{su2embed}
	r=Q^{-1}\bar{r}\qquad \omega_j(r)\equiv w_j^{1/2}\omega(\bar{r}),\qquad a_i\equiv \frac{a(\bar{r})}{2},\qquad m\equiv Q\tilde{m}(\bar{r}),\qquad \ell\equiv Q\tilde{\ell},
\end{equation}
for all $i\in\{1,...,\mc{L}\}$, with $A(r)$ and $W_+(r)$ set in the basis \eqref{Wplus}, and with $Q^2$ given in \eqref{Qdef}, noting that \eqref{lamdef} implies $\sum_{j=1}^\mc{L}w_j=Q^2$. Then the field equations become
\begin{equation}
	\begin{split}
		\tilde{m}'&=\frac{\bar{r}^2}{2S^2}\left(\frac{d a }{d\bar{r}}\right)^{\!\!2}+\frac{ a ^2\omega^2}{\mu S^2}+\mu\left(\frac{d\omega}{d\bar{r}}\right)^{\!\!2}+\frac{(k-\omega^2)^2}{2\bar{r}^2},\\
		\frac{1}{S}\frac{dS}{d\bar{r}}&=\frac{2}{\bar{r}}\left(\frac{d\omega}{d\bar{r}}\right)^{\!\!2}+\frac{2\omega^2 a ^2}{\bar{r}\mu^2 S^2},\\
		0&=\bar{r}^2\mu\frac{d^2\omega}{d\bar{r}^2}+\left(2\tilde{m}-\frac{(k-\omega^2)^2}{\bar{r}}+\frac{\bar{r}^3}{\tilde{\ell}^2}-\frac{\bar{r}^3}{2S^2}\left(\frac{da}{d\bar{r}}\right)^{\!\!2}\right)\frac{d\omega}{d\bar{r}}+\frac{\bar{r}^2 a ^2\omega}{\mu S^2}+\omega(k-\omega^2),\\
		0&=\bar{r}^2\mu\frac{d^2 a}{d\bar{r}^2}+\bar{r}^2\mu\left(\frac{2}{r}-\frac{S^\prime}{S}\right)\frac{da}{d\bar{r}}-2a\omega^2,
	\end{split}
\end{equation}
with
	\labeq{\mu(\bar{r})=k-\frac{2\tilde{m}}{\bar{r}}+\frac{\bar{r}^2}{\tilde{\ell}^2}.}{}
These equations are identical to those for the dyonic $\essu$ adS case for $k=1$, and for general $k$ and $a=0$, we obtain the purely magnetic topological $\essu$ equations. The existence of (nodeless) solutions has been proven in both of these cases \cite{nolan_existence_2012,van_der_bij_new_2002}.

\section{Proof of local existence at the boundaries $r=0$, $r=r_h$, $r\rar\infty$}\label{sec:LocEx}

Since we know the boundary conditions to expect, we can turn our attention to proving the local existence of solutions near those boundaries. To do this, we rely on a well-known theorem of differential equations \cite{breitenlohner_static_1994}, generalised to the appropriate case by \cite{oliynyk_local_2002}.
\begin{thr} \cite{oliynyk_local_2002}\label{Kthm3}
	The system of differential equations
	\begin{equation}
	\begin{split}
	t\frac{du_i}{dt}&=t^{\mu_i}f_i(t,u,v),\\
	t\frac{dv_i}{dt}&=-h_i(u)v_i+t^{\nu_i}g_i(t,u,v)\\
	\end{split}
	\end{equation}
	where $\mu_i,\,\nu_i\in\Z_{>1}$, $f_i$, $g_i$ are analytic functions in a neighbourhood of $(0,c_0,0)\in\R^{1+m+n}$, and the functions $h_i\,:\,\R^m\rar\R$ are positive in a neighbourhood of $c_0\in\R^m$, has a unique solution $t\mapsto(u_i(t),v_i(t))$ such that
	\begin{equation}
	u_i(t)=c_i+O(t^{\mu_i}),\qquad\mbox{ and }v_i(t)=O(t^{\nu_i}),
	\end{equation}
	for $|t|>\bar{r}$ for some $\bar{r}>0$ if $|c-c_0|$ is small enough. Moreover, the solution depends analytically on the parameters $c_i$.
\end{thr}
The proof of this theorem involves an approach similar to the proof of the Picard-Lindelh\"{o}f existence theorem for initial value problems \cite{coddington_theory_1955}. Now, for each of the boundaries $r=0$, $r=r_h$ and $r\rar\infty$, we proceed by first identifying the boundary conditions that we expect, and then by formulating the field equations in a form such that we may apply Theorem \ref{Kthm3}.

\subsection{Local existence at the origin}

\subsubsection{Boundary Conditions at $r=0$}

Near the origin $r=0$ we may simply use the independent variable $r$, and hence we expand functions like $f(r)=\sum_{k=0}^\infty f_kr^k$. Thus we obtain the following recurrence relations. The Einstein equations give for $m_{k+1}$, $S_{k}$:
\begin{subequations}\label{recr0}
	\begin{align}
	(k+1)m_{k+1}=& G_k+\frac{1}{\ell^2}G_{k-2}+P_k+\frac{\eta_{k-4}}{4\ell^2S_0^2}+\frac{\zeta_{k}}{4S_0^2}+\frac{\eta_{k-2}}{4S_0^2}+\sum\limits_{l=2}^{k-2}\bar{M}_l,\\
	kS_k=& 2G_k+\frac{\zeta_k}{2S_0^2}+\sum\limits_{l=2}^{k-2}\bar{S}_l,
	\end{align}
\end{subequations}
and the Yang-Mills equations give for $\omega_{k+1}$ and $\mc{E}_k$:
\begin{subequations}
	\begin{align}
	b_{i,k+1}&=\slim_{j=1}^\mc{L}\left(T_{ij}-k(k+1)\bsym{\delta}_{ij}\right)\omega_{j,k+1},\label{orb}\\
	z_{i,k}&=\slim_{j=1}^\mc{L}\left(T_{ij}-k(k+1)\bsym{\delta}_{ij}\right)\mc{E}_{j,k}.\label{orz}
	\end{align}
\end{subequations}
Here, $\bv{T}\equiv T_{ij}$ is the matrix defined by
\labeq{T_{ij}\equiv\omega_{i,0}C_{ij}\omega_{j,0},}{Adef}
$\bsym{\delta}_{ij}$ is the Kronecker symbol, the left-hand side of \eqref{orb}, \eqref{orz} are the vectors $\bv{b}_k\equiv(b_{1,k},...,b_{\mc{L},k})$, and $\bv{z}_k$ (defined similarly). The quantities $\bv{b}_k$,  $\bv{z}_k$, $\bar{\bv{S}}_k$ and $\bar{\bv{M}}_k$ are complicated expressions whose form is unimportant here. 

We can see that these equations are identical to the dyonic $\sun$ case \cite{baxter_general_2016}, and so as in that case, we may solve \eqref{EEs} and \eqref{YMEs} near $r=0$ and obtain a solution with $2\mc{L}$ free parameters on condition that the recurrence relations \eqref{orb} and \eqref{orz} can be solved. This in turn is conditional upon the vectors $\bv{b}_k$ and $\bv{z}_k$ lying in the left kernel of the matrix $\bv{T}$. Particular methods will exist for this purpose in each Lie group: in Section \ref{sec:sl2c}, we state and generalise proofs in \cite{oliynyk_local_2002} which depend directly on the root structure of the Lie algebra $\mk{g}$ treated as an $\mk{sl}(2,\C)$ submodule.

We note here that $G_k=P_k=\zeta=0$ for $k<2$. For the lower order terms, we find:
\begin{equation}
	S_0\neq 0,\quad m_0=m_1=m_2=0,\quad \mc{E}_{j,0}=0,\quad\omega^2_{j,0}=w_j,\quad \omega_{j,1}=0.
\end{equation}
%

\begin{table}\label{spectable}
	\centering
	\begin{tabular}{ || c | c || }
		\hline
		\hline
		Lie algebra & $\mc{E}$\\
		\hline
		\hline
		$A_\mc{L}$ & $j$\\
		\hline
		$B_\mc{L}$ & $2j-1$\\
		\hline
		$C_\mc{L}$ & $2j-1$\\
		\hline
		$D_\mc{L}$ & 
		$\Bigg\{\begin{array}{ll}
		2j-1 & \mbox{ if }j\leq\,(\mc{L}+2)/2\\
		\mc{L}-1 & \mbox{ if }j=(\mc{L}+2)/2\\
		2j-3 & \mbox{ if }j\mbox{ \textgreater }\,(\mc{L}+2)/2\\
		\end{array}$\\
		\hline
		$G_2$ & 1, 5\\
		$F_4$ & 1, 5, 7, 11\\
		$E_6$ & 1, 4, 5, 7, 8, 11 \\
		$E_7$ & 1, 5, 7, 9, 11, 13, 17\\
		$E_8$ & 1, 7, 11, 13, 17, 19, 23, 29\\
		\hline
		\hline
	\end{tabular}
	\caption{This table shows $\mc{E}$, for the calculation of $\mbox{spec}(\bv{T})=\{k(k+1)\,|\,k\in\mc{E}\}$. For the classical Lie algebras the table shows $k_j$ for $j=1,...,\mc{L}$, $\mc{L}=\mbox{rank}(\mk{g})$. We note that $1\in\mc{E}$ always so that $k=1$ belongs to all Lie algebras.}
	\label{Table1}
	\centering
\end{table}

The higher order coefficients which remain arbitrary are at the orders $r^k$ for which $k(k+1)$ is an eigenvalue of the matrix $\bv{T}$. But the eigenvalues of $\bv{T}$ are $k(k+1)$ for a set $\{k\}$ depending on the Lie algebra in question. For all the simple Lie algebras, we may calculate the spectrum of eigenvalues from the Cartan matrix by using the definition \eqref{Adef}; see Table 1 for this information. 
The proof for the classical Lie algebras then follows from the root structure, i.e. the results in Section \ref{sec:sl2c}.

We will see in Section \ref{sec:locExOr} that in some neighbourhood of $r=0$, the relevant field variables have the following behaviour:
\begin{equation}\label{BCorigin}
	\begin{split}
		m(r)&=m_3r^3+O(r^4),\\
		S(r)&=S_0+O(r^2),\\
		\omega_i(r)&=\omega_{i,0}+\slim_{j=1}^\mc{L} Q_{ij}\hat{u}_j(r)r^{\kappa_j+1},\\
		\mc{E}_i(r)&=\slim_{j=1}^\mc{L} Q_{ij}\hat{\psi}_j(r)r^{\kappa_j},\quad (i=1,...,\mc{L}).
	\end{split}
\end{equation}
Here, $Q_{ij}$ is a non-singular matrix, $\kappa_j$ are integers and $\hat{u}_j$ and $\hat{\psi}_j$ are some functions which will be defined in Section \ref{sec:locExOr}. Also, $m_3$ is a constant fixed by the gauge functions, $S_0$ is fixed by the requirement that $S\rar1$ as $r\rar\infty$, and $\omega^2_{j,0}=w_j$. Therefore we have $2\mc{L}$ free solution parameters here in total.

\subsubsection{Necessary results for local existence at $r=0$}\label{sec:sl2c}

Now we are ready to state a series of results proven in \cite{oliynyk_local_2002} which will help us to prove existence locally at $r=0$. Essentially, these are necessary because we find that the terms $\mc{F}$ and $\mc{Z}$ in the Yang-Mills equations \eqref{YMEs} are troublesome in that in general they contain non-regular terms. The results of this Section are necessary to ensure that these terms are identically zero. We emphasise that solitons are only available for the spherical case $k=1$; hence for the rest of this section, $k\in\Z$ is simply used as an index.

First we introduce our conventions. We begin by defining a non-degenerate Hermitian inner product $\HIP{\,}{\,}:\mk{g}\times\mk{g}\rar\C$, such that
\begin{equation}
	\langle X\,|\,Y\rangle\equiv-(c(X),Y)\quad\forall\,X,Y\in\mk{g}.
\end{equation}
Then $\HIP{\,}{\,}$ is a real positive definite inner product on $\mk{g}_0$, since $c:\mk{g}\rar\mk{g}$ is the conjugation operator determined on the compact real form $\mk{g}_0$. We may show that $\HIP{\,}{\,}$ satisfies
\begin{equation}
	\begin{split}
		\langle X\,|\,Y\rangle&=\overline{\langle Y\,|\,X\rangle},\\
		\HIP{c(X)}{c(Y)}&=\overline{\langle X\,|\,Y\rangle},\\
		\HIP{[X,c(Y)]}{Z}&=\HIP{X}{[Y,Z]}
	\end{split}
\end{equation}
for all $X,Y,Z\in\mk{g}$. We use this to create a positive definite, real inner product $\HHIP{\,}{\,}:\mk{g}\times\mk{g}\rar\R$, with
\begin{equation}\label{HHIP}
	\HHIP{X}{Y}\equiv\mbox{Re}\HIP{X}{Y}\quad\forall X,Y\in\mk{g}.
\end{equation}
Let $\|\,\,\|$ be the norm induced by \eqref{HHIP}, i.e. $\|X\|^2=\HHIP{X}{X}\,\forall X\in\mk{g}$. Then we may verify the following properties of $\HHIP{\,}{\,}$ for all $X,Y,Z\in\mk{g}$:
\begin{equation}
	\begin{split}
		\HHIP{X}{Y}&=\HHIP{Y}{X},\\
		\HHIP{c(X)}{c(Y)}&=\HHIP{X}{Y},\\
		\HHIP{[X,c(Y)]}{Z}&=\HHIP{X}{[Y,Z]}.
	\end{split}
\end{equation}
Let $\Omega_+,\Omega_-\in\mk{g}$ be two vectors such that
\begin{equation}
	[W_0,\Omega_{\pm}]=\pm 2\Omega_\pm,\quad\quad [\Omega_+,\Omega_-]=W_0,\quad\quad c(\Omega_+)=-\Omega_-.
\end{equation}
Then $\mbox{span}_\C\{W_0,\Omega_+,\Omega_-\}\cong\mk{sl}(2,\C)$. We again use a central dot notation $\cdot$ to represent the adjoint action:
\begin{equation}
	X\ccdot Y\equiv\mbox{ad}(X)(Y),\qquad\forall X\in\mbox{span}_\C\{W_0,\Omega_+,\Omega_-\},\,Y\in\mk{g}.
\end{equation}
Since $W_0$ is a semisimple element, $\mbox{ad}(W_0)$ is diagonalisable, and so basic $\mk{sl}(2)$ representation theory says that we know that the eigenvalues must be integers. Therefore we define $V_n$ as the eigenspaces of $\mbox{ad}(W_0)$, i.e. with
\begin{equation}
	V_n\equiv\{X\in\mk{g}\,|\,W_0\ccdot X=nX,\,n\in\Z\,\}.
\end{equation}
Given that we are investigating the principal case here, it follows that $V_2$ is the space we are most interested in. It also follows from $\mk{sl}(2,\C)$ representation theory that if $X\in\mk{g}$ is a highest weight vector of the adjoint representation of $\mbox{span}_\C\{W_0,\Omega_+,\Omega_-\}$ with weight $n$, and we define $X_{-1}=0$, $X_0=X$ and $X_j=(1/j!)\Omega_-^j\ccdot X_0$ ($j\geq 0$), then
\begin{equation}\label{ladop}
	\begin{split}
		W_0\ccdot X_j & =(n-2)X_j,\\
		\Omega_-\ccdot X_j & =(j+1)X_{j+1},\\
		\Omega_+\ccdot X_j & =(n-j+1)X_{j-1},
	\end{split}
\end{equation}
so that $\Omega_\pm$ act as `ladder' operators, taking us between the various weight spaces. 
\begin{prop}\cite{oliynyk_local_2002}\label{Kprop1}
	There exist $\bsym{\Sigma}$ highest weight vectors $\xi^1$, $\xi^2,$,... $\xi^{\bsym{\Sigma}}$ for the adjoint representation of  $\emph{span}_\C\{W_0,\Omega_+,\Omega_-\}$ on $\mk{g}$ that satisfy
	\renewcommand{\labelenumii}{\Roman{enumii}}
	\begin{enumerate}
		\item the $\xi^j$ have weights $2k_j$ where $j=1,...,\bsym{\Sigma}$ and $1=k_1\leq k_2\leq...\leq k_{\bsym{\Sigma}}$;\\
		\item if $V(\xi^j)$ denotes the irreducible submodule of $\mk{g}$ generated by $\xi^j$, then the sum $\slim_{j=1}^{\bsym{\Sigma}}V(\xi^j)$ is direct;\\
		\item if $\xi^j_l=(1/l!)\Omega_-^l\ccdot\xi^j$, then $c(\xi^j_l)=(-1)^l\xi^j_{2k_j-l}$;\\
		\item $\bsym{\Sigma}=|\Sigma_\lambda|$ and the set $\{\xi^j_{k_j-1}\,|\,j=1,...,\bsym{\Sigma}\}$ forms a basis for $V_2$ over $\C$.\\
	\end{enumerate}
\end{prop}
This proposition establishes a basis over $V_2$ of weight vectors $\xi_j$, and their properties in terms of the operators \eqref{ladop}. To deal with the problem terms in the gauge equations, we define an $R$-linear operator $T:\mk{g}\rar\mk{g}$ by
\begin{equation}\label{Tdef}
	T\equiv\frac{1}{2}\mbox{ad}(\Omega_+)\circ\left(\mbox{ad}(\Omega_-)+\mbox{ad}(\Omega_+)\circ c\right).
\end{equation}
It is proven in \cite{oliynyk_local_2002} that $T$ \emph{a)} is symmetric with respect to the inner product $\HHIP{\,}{\,}$, i.e. $\HHIP{T(X)}{Y}=\HHIP{X}{T(Y)}\,\,\forall X,Y\in\mk{g}$, and \emph{b)} restricts to $V_2$, i.e. $T(V_2)\subset V_2$, therefore we define $T_2\equiv T|_{V_2}$.

Now we label the set of integers $k_j$ from Proposition \ref{Kprop1} as follows:
\begin{equation}
	\begin{split}
		1=k_{J_1}=k_{J_1+1}=...=k_{J_1+k_1-1}&< k_{J_2}=k_{J_2+1}=...=k_{J_2+m_2-1}\\
		&<...\\
		&<k_{J_I}=k_{J_I+1}=...=k_{J_I+m_I-1},\\
	\end{split}
\end{equation}
where we define the series of integers $J_1=1$, $J_k+m_k=J_{k+1}$ for $k=1,...,I$ and $J_{I+1}=\bsym{\Sigma}-1$. To ease notation we define
\begin{equation}\label{kapdef}
	\kappa_j\equiv k_{J_j},\mbox{   for }j=1,...,I.
\end{equation}

As noted in Proposition \ref{Kprop1}, the set $\{\xi^j_{k_j-1}\,|\,j=1,...,\bsym{\Sigma}\}$ forms a basis of $V_2$ over $\C$. Therefore the set of vectors $\{X^l_s,Y^l_s\,|\,l=1,...,I;\,s=0,1,...,m_l-1\}$, where $X^l_s=\xi^{J_l+s}_{\kappa_l-1}$ for $\kappa_l$ odd, or $X^l_s=i\xi^{J_l+s}_{\kappa_l-1}$ for $\kappa_l$ even, and $Y^l_s=iX^l_s$. Then $T$ is symmetric, and so also is $T_2$, and hence $T_2$ must be diagonalizable. Then the following Lemma is true.
\begin{lem}\label{Klem2}
	\begin{equation}
	T_2(X^l_s)=\kappa_l(\kappa_l+1)X^l_s\qquad\mbox{and}\qquad T_2(Y^l_s)=0,
	\end{equation}
	for $l=1,...,I$ and $s=0,1,...,m_l-1$.
\end{lem}
In other words, the set $\{X^l_s,Y^l_s\,|\,l=1,...,I;\,s=0,1,...,m_l-1\}$ forms an eigenbasis of $T_2$. An immediate consequence of this is that $\mbox{spec}(T_2)=\{0\}\cup\{\kappa_j(\kappa_j+1)\,|\,j=1,...,I\}$, and $m_j$ is the dimension of the eigenspace associated to the eigenvalue $\kappa_j(\kappa_j+1)$ ($I$ being the number of distinct positive eigenvalues of $T_2$).

Following Lemma \ref{Klem2}, we define the spaces
\begin{equation}\label{E_0q}
	E^l_0\equiv\mbox{span}_\R\{Y^l_s\,|\,s=0,1,...,m_l-1\},\quad E^l_+\equiv\mbox{span}_\R\{X^l_s\,|\,s=0,1,...,m_l-1\},
\end{equation}
and
\begin{equation}\label{E_0}
	E_0\equiv\bigoplus_{l=1}^{I}E^l_0,\qquad E_+\equiv\bigoplus_{l=1}^{I}E^l_+.
\end{equation}
Then $E_0=\mbox{ker}(T_2)$ and $E^j_+$ is the eigenspace of $T_2$ corresponding to the eigenvalue $\kappa_j(\kappa_j+1)$. Also, from Proposition \ref{Kprop1} (iv) we see that $V_2=E_0\oplus E_+$.

\begin{lem}\label{Klem3}
	Suppose $X\in V_2$. Then $X\in\bigoplus_{q=1}^lE^q_0\oplus E_+^q$ if and only if $\Omega^{\kappa_l}_+\ccdot X=0$.
\end{lem}
\begin{lem}\label{Klem4}
	Suppose $X\in V_2$. Then $X\in\bigoplus_{q=1}^lE^q_0\oplus E^q_+$ if and only if $\Omega_+^{\kappa_l+2}\ccdot c(X)=0$.
\end{lem}
\begin{lem}\label{Klem5}
	Let $\,\widetilde{\,\,}\,:\,\Z_{\geq -1}\rar\{1,2,...,I\}$ be the map defined by
	\begin{equation}
		\widetilde{-1}=\tilde{0}=1\quad\mbox{ and }\quad\tilde{s}=\max\,\{l\,|\,\kappa_l\leq s\}\mbox{ if }s>0.
	\end{equation}
	Then
	\begin{enumerate}
		\item $\kappa_{\tilde{s}}\leq s$ for every $s\in\Z_{\geq0}$,\\
		\item $\kappa_{\tilde{s}}\leq s\leq\kappa_{\tilde{s}+1}$ for every $s\in\{0,1,...,\kappa_{I-1}\}$.
	\end{enumerate}
\end{lem}
\begin{lem}\label{Klem6}
	If $X\in V_2$, $\kappa_{\tilde{p}}+s<\kappa_{\tilde{p}+1}\,(s\geq 0),$ and $\Omega_+^{\kappa_{\tilde{p}}+s}\ccdot X=0$, then $\Omega^{\kappa_{\tilde{p}}}_{+}\ccdot X=0$.
\end{lem}

The next two theorems are the crux of the proof of local existence at the origin. The first was proven in \cite{oliynyk_local_2002}, using the above Lemmata \ref{Klem3} to \ref{Klem6}:
\begin{thr}\label{Kthr9}
	Suppose $p\in\{1,2,...,\kappa_I-1\}$ and let $Z_{0},Z_{1},...,Z_{p+1}\in V_2$ be a sequence of vectors satisfying $Z_{0}\in E^1_0\oplus E^1_+$ and $Z_{n+1}\in\bigoplus_{q=1}^{\tilde{n}}E^q_0\oplus E^q_+$ for $n=0,1,...,p$. Then for every $j\in\{1,2,...,p+1\}$, $s\in\{0,1,...,j\}$,
	\begin{enumerate}
		\item $[[c(Z_{j-s}),Z_{s}],Z_{p+2-j}]\in\bigoplus_{q=1}^{\tilde{p}}E^q_0\oplus E^q_+$, \\
		\item $[[c(Z_{p+2-j}),Z_{j-s}],Z_{s}]\in\bigoplus_{q=1}^{\tilde{p}}E^q_0\oplus E^q_+$.
	\end{enumerate}
\end{thr}
The second is an extension of Theorem \ref{Kthr9} that we must make, in order to include the electric gauge field. It is very similar, but we must use two sequences of vectors here.
\begin{thr}\label{MyThrr0}
	Suppose $s\in\{1,2,...,\kappa_I-1\}$ and let $Z_{i,0},Z_{i,1},...,Z_{i,s+1}\in V_2$ (for $i\in\{1,2\}$) be two sequences of vectors satisfying $Z_{i,0}\in E^1_0\oplus E^1_+$ and $Z_{i,n+1}\in\bigoplus_{q=1}^{\tilde{n}}E^q_0\oplus E^q_+$ for $n=0,1,...,s$. Then for every $l\in\{1,2,...,s+1\}$, $m\in\{0,1,...,l\}$, the following terms all lie in $\bigoplus_{q=1}^{\tilde{s}}E^q_0\oplus E^q_+$:
	\begin{equation}
		\begin{split}
			(i)&\,\,[[c(Z_{2,s-j+2}),Z_{1,j}],Z_{2,0}],\qquad (ii)\,\,[[c(Z_{2,0}),Z_{1,j}],Z_{2,s-j+2}],\\
			(iii)&\,\,[[c(Z_{2,0}),Z_{2,s-j+2}],Z_{1,j}]\qquad
			(iv)\,\,[[c(Z_{1,j}),Z_{2,q-j+1}],Z_{2,s-q+1}].\\
		\end{split}
	\end{equation}
\end{thr}
\textbf{Proof} We demonstrate the proof using point \emph{(iv)}; the others are similar and simpler. The thrust of it is that all of the results in this Section so far will apply to both sequences of vectors $Z_{1,k}$ and $Z_{2,k}$. Hence, we can use Lemmata \ref{Klem3} and \ref{Klem4} to show that
\begin{equation}\label{Om0}
	\Omega_+^{\kappa_{(n-1)^{\tilde{\,}}}}.Z_{i,n}=\Omega_+^{\kappa_{(n-1)^{\tilde{\,}}}+2}.c(Z_{i,n})=0
\end{equation}
for $i\in\{1,2\}$ and $n\in\{0,1,...,s+1\}$. Also, if $l\in\{1,2,...,s+1\}$, $m\in\{0,1,...,l\}$, we find
\begin{equation}
	\Omega_+^s.[[c(Z_{1,j}),Z_{2,q-j+1}],Z_{2,s-q+1}]=\slim_{l=0}^s\slim_{m=0}^l
	\begin{pmatrix}
		s\\l
	\end{pmatrix}
	\begin{pmatrix}
		l\\m
	\end{pmatrix}
	a_{jlmqs}
\end{equation}
with
\begin{equation}
	a_{jlmqs}=[[\Omega_+^{m}.c(Z_{1,j}),\Omega_+^{l-m}.Z_{2,q-j+1}],\Omega_+^{s-l}.Z_{2,s-q+1}].
\end{equation}
Now we apply \eqref{Om0}, implying that $a_{jlmqs}=0$ if $m-2\geq\kappa_{(j-1)^{\tilde{\,}}}$ or $l-m\geq\kappa_{(q-j)^{\tilde{\,}}}$ or $s-l\geq\kappa_{(s-q)^{\tilde{\,}}}$ --  by Lemma \ref{Klem5}, this becomes the condition $m-2\geq j-1$ or $l-m\geq q-j$ or $s-l\geq s-q$. Therefore, $a_{jlmqs}\neq0$ only if we can find integers $m$, $l$ such that $q<l<q+m-j<q+1$, which is impossible. Hence $a_{jlmqs}=0$ for all $l$, $m$, and we find that $\Omega_+^s.[[c(Z_{1,j}),Z_{2,q-j+1}],Z_{2,s-q+1}]=0$. From Lemmata \ref{Klem5} and \ref{Klem6}, this implies that $\Omega_+^{\tilde{s}}.[[c(Z_{1,j}),Z_{2,q-j+1}],Z_{2,s-q+1}]=0$, and hence $[[c(Z_{1,j}),Z_{2,q-j+1}],Z_{2,s-q+1}]\in\bigoplus_{q=1}^{\tilde{s}}E^q_0\oplus E^q_+$. The other three terms follow very similarly. $\Box$
\begin{prop}\label{Kprop4}
	Let $W_0$ be regular. Then if $\Omega_+\in\slim_{\alpha\in \Sigma_\lambda}\R\bv{\emph{e}}_{\alpha}$, $E_+=\slim_{\alpha\in \Sigma_\lambda}\R\bv{\emph{e}}_{\alpha}$.
\end{prop}
This is sufficient to establish that
\begin{equation}
	T_2(\bv{e}_\alpha)=\slim_{\beta\in\Sigma}\omega_\alpha\langle\alpha,\beta\rangle\omega_\beta,
\end{equation}
which in our basis becomes
\begin{equation}
	T_2(\bv{e}_\alpha)=\slim_{\beta\in\Sigma}\omega_\alpha C_{\alpha\beta}\omega_\beta\bv{e}_\beta,
\end{equation}
which ties in with \eqref{Adef}.

\subsubsection{Proof of local existence at the origin ($r=0$)}\label{sec:locExOr}

In this Section we use Theorem \ref{Kthm3} and the results of Section \ref{sec:sl2c} to prove the existence of solutions, unique and analytic with respect to their boundary parameters, in some neighbourhood of the origin. To do this, we find it necessary first to rewrite the electric gauge equation \eqref{YME} in a new form. We introduce
\begin{equation}
	\Omega_\pm\equiv W_\pm(0)=\slim_{j=1}^\mc{L}w^{1/2}_j\bv{e}_{\pm\alpha_j}.
\end{equation}
Then, we define
\begin{equation}\label{defEhat}
	\hat{E}_{\pm}\equiv-i[A,\Omega_\pm].
\end{equation}
From this, and using $c(A)=A$ and $c(\Omega_+)=-\Omega_-$, we may easily derive that $c(\hat{E}_+)=\hat{E}_-$. By repeated use of the Jacobi identity, and noting that commutators between combinations of $W_\pm$, $\Omega_\pm$ are either zero or lie in $\mk{h}$, we may see that
\begin{equation}\label{WAW}
	[[W_+,[A,W_-]],\Omega_+]\equiv[[W_+,[A,\Omega_+]],W_-];
\end{equation}
and thus taking the commutator of \eqref{YMAComm} with $\Omega_+$, the electric gauge equation may be rewritten as
\begin{equation}\label{YMEComm}
	\mu S\left(\frac{r^2\hat{E}'_+}{S}\right)^\prime=[W_+,[\hat{E}_+,W_-]].
\end{equation}
It is worthwhile noting a few things. Firstly, that if we write $\hat{E}_\pm$ in our familiar basis \eqref{wBasis}, e.g. $\hat{E}_\pm=\sum_{\alpha\in\Sigma}\hat{\mc{E}}_\alpha\bv{e}_{\pm\alpha}$, then these are related to the earlier functions $\mc{E}_\alpha$ from \eqref{Ebar} by $\hat{\mc{E}}_\alpha=\mc{E}_\alpha w^{1/2}_\alpha$, so that we have overall gained at most a constant factor on each electric gauge function, and this will in any case be removed again later. Secondly, using (\ref{Tdef}, \ref{defEhat}) and the Jacobi identity, it is clear that
\begin{equation}\label{A2Ehat}
	T_2(\hat{E}_+)=-[\Omega_+,[\hat{E}_+,\Omega_-]].
\end{equation}
Finally, using Proposition \ref{Kprop4}; from previous results \cite{oliynyk_local_2002}; and noting that the proof of Proposition 4 carries over to show that $\hat{E}_+(r)\in E_+$; we see that the solutions $W_+(r)$, $\hat{E}_+(r)$ of equations (\ref{YMWComm}, \ref{YMEComm}), are completely characterised by the condition
\begin{equation}\label{InEPlus}
	W_+(r),\hat{E}_+(r)\in E_+\,\,\,(\forall r).
\end{equation}
Before embarking upon our proof, we also state some definitions. First, we define the set of integers defining the eigenvalues of $T_2$ as $\mc{E}$:
\begin{equation}
	\mc{E}\equiv\{\kappa_j\,|\,j=1,...,I\},
\end{equation}
for $\kappa_j$ given in \eqref{kapdef}. Given \eqref{InEPlus}, we introduce new functions $u_{i}(r),\psi_i(r)$ with
\begin{equation}\label{K1}
	W_+(r)=\Omega_++\sum_{s\in\mc{E}}u_{s+1}(r)r^{s+1},\qquad\hat{E}_+(r)=\sum_{s\in\mc{E}}\psi_{s}(r)r^{s}
\end{equation}
and with $\Omega_\pm=W_\pm(0)$ and $\psi_s(r),u_{s+1}(r)\in E_+^{\tilde{s}}\,\forall r,\,\forall s\in\mc{E}$. Since $E_+=\bigoplus_{q=1}^IE_+^q$, these transformations are clearly invertible. Finally we define a symbol $\chi$ which will pick out the required orders of terms:
\begin{equation}
	\chi_{s+1}=\Bigg\{
	\begin{array}{ll}
		1 & \mbox{ if }s\in\mc{E},\\
		0 & \mbox{ otherwise.}\\
		\end{array}
\end{equation}
Now we state the proposition.

\begin{prop}\label{prop:lex0}
	In a neighbourhood of the origin $r=0$ for solitons only, there exist regular solutions to the field equations, analytic and unique with respect to their initial values, of the form
	\begin{equation}\label{prop0exp}
		\begin{split}
			m(r)&=m_3r^3+O(r^4),\\
			S(r)&=S_0+O(r^2),\\
			\omega_i(r)&=\omega_{i,0}+\slim_{j=1}^\mc{L} Q_{ij}\hat{u}_j(r)r^{\kappa_j+1},\\
			\mc{E}_i(r)&=\slim_{j=1}^\mc{L} Q_{ij}\hat{\psi}_j(r)r^{\kappa_j},\quad (i=1,...,\mc{L}).\\
		\end{split}
	\end{equation}
	Above, $Q_{ij}$ is a non-singular matrix for which the $j$th column is the eigenvector of the matrix $\emph{\bv{T}}$ \eqref{Adef} with eigenvalue $\kappa_j(\kappa_j+1)$, and $\hat{u}_j(r)$, $\hat{\psi}_j(r)$ are some functions of $r$. Each solution is entirely and uniquely determined by the initial values $\hat{u}_j(0)\equiv\tilde{u}_j$ and $\hat{\psi}_j(0)\equiv\tilde{\psi}_j$, for $\tilde{u}_j$, $\tilde{\psi}_j$ arbitrary. Once these are determined, the metric functions $m(r)$ and $S(r)$ are entirely determined.
\end{prop}
\textbf{Proof} 

Using the above definitions, we may write \eqref{K1} as 
\begin{equation}\label{wandE}
	W_+(r)\equiv\Omega_++U_+=\Omega_++\slim_{i=1}^{\infty}\chi_iu_i(r)r^i,\qquad \hat{E}_+(r)=\slim_{i=1}^{\infty}\chi_{i+1}\psi_i(r)r^i.
\end{equation}
It is also sometimes more convenient to represent the electric field in a basis more like the original basis:
\begin{equation}\label{defAhat}
	A=\slim_{s\in\mc{S}}\hat{a}_sr^s=\slim_{i=1}^\infty\chi_{i+1}\hat{a}_{i}r^i\in E_0.
\end{equation}
It is clear from this definition that $\hat{a}_p\in E^{\tilde{p}}_0$ for all $p\in\mc{S}$; and since $E_0=\bigoplus_{q=1}^{I}E_0^q$, this transform of $A$ is invertible. Substituting \eqref{wandE} into the Yang-Mills equations \eqref{YMEComm} and \eqref{YMWComm} and using $T_2(u_{s+1})=s(s+1)u_{s+1}$ and $T_2(\psi_{s})=s(s+1)\psi_{s}$, we find that:
\begin{equation}\label{BigF}
	\mc{F}=-\slim_{s\in\mc{E}}s(s+1)r^{s+1}+\slim_{s=2}^{N_1}f_sr^s,\qquad\mc{Z}=\slim_{s\in\mc{E}}s(s+1)r^{s}+\slim_{s=3}^{N_2}g_sr^s
\end{equation}
for some $N_1,N_2\in\mathbb{Z}$, and with
\begin{equation}\label{fdef}
	\begin{split}
		f_s=&\frac{1}{2}\slim_{j=2}^{s-2}\bigg\{\left[\left[\Omega_+,c(\chi_ju_j)\right]+\left[\Omega_-,\chi_ju_j\right],\chi_{s-j}u_{s-j}\right]\\
		&+\left[\left[\chi_ju_j,c(\chi_{s-j}u_{s-j})\right],\Omega_+\right]+\slim_{m=2}^{j-2}\left[\left[\chi_m u_m,c(\chi_{j-m}u_{j-m})\right],\chi_{s-j}u_{s-j}\right]\bigg\},\\
		g_s=&-\frac{1}{2}\slim_{j=1}^{s-2}\bigg\{\left[\Omega_+,\left[c(\chi_{s-j}u_{s-j}),\chi_{j+1}\psi_j\right]\right]+\left[\Omega_-,\left[\chi_{s-j}u_{s-j},\chi_{j+1}\psi_j\right]\right]\bigg\}\\
		&-\slim_{j=1}^{s-3}\slim_{m=2}^{j-2}\left[\chi_{s-m-1}u_{s-m-1},\left[\chi_{m-j+1}u_{m-j+1},c(\chi_{j+1}\psi_j)\right]\right].
	\end{split}
\end{equation}
We define new variables $v_{s+1}\equiv u'_{s+1},\,q_s=\psi^\prime_s,\,\,\forall s\in\mc{E}$ and we introduce an analytic map $\hat{C}:E_0\times E_+\rar E_+$ by $\hat{C}(A,W_+)=[A,[A,W_+]]$, which in our bases \eqref{wandE}, \eqref{defAhat}, becomes 
\begin{equation}
	\begin{split}
		\hat{C}&=-\slim_{k\in\mc{E}}\left(\sum_{i=1}^\infty\left([\chi_{i+1}\hat{a}_i,\psi_{k}+\slim_{j=1}^{i-1}[\chi_{i-j+1}\hat{a}_{i-j},u_{k+1}]]\right)r^{i-1}\right)r^{k+1}\\
		&\equiv\slim_{k\in\mc{E}}\hat{C}_kr^{k+1}.
	\end{split}
\end{equation}
Then the Yang-Mills equations can be written
\begin{equation}\label{slimkinE}
	\begin{split}
		r\slim_{k\in\mc{E}}v'_{k+1}r^{k+1}=&-2\slim_{k\in\mc{E}}(k+1)v_{k+1}r^{k+1}+\slim_{k\in\mc{E}}\frac{k(k+1)}{r}\left(\frac{1}{\mu}-1\right)u_{k+1}r^{k+1}\\
		&-\frac{2}{r\mu}\left(m-\frac{P}{r}+\frac{r^3}{\ell^2}-\frac{r^3\eta}{\mu S^2}\right)\slim_{k\in\mc{E}}\left(v_{k+1}r^{k+1}+(k+1)u_{k+1}r^{k+1}\right)\\
		&-\frac{r^2}{\mu S^2}\slim_{k\in\mc{E}}\hat{C}_{k}r^{k+1}-\frac{1}{\mu}\slim_{k=2}^{N_1}f_kr^{k-1},\\
		r\slim_{k\in\mc{E}}q'_kr^k=&-2\slim_{k\in\mc{E}}(k+1)q_kr^k+\left(\frac{1}{\mu}-1\right)\slim_{k\in\mc{E}}k(k+1)r^{k-1}\psi_k\\
		&-\frac{2}{r}\left(G+\frac{\zeta}{\mu^2 S^2}\right)\slim_{k\in\mc{E}}\left(q_kr^{k+1}+k\psi_kr^k\right)+\frac{1}{\mu}\slim_{j=3}^{N_2}g_{j}r^{j-1}.
	\end{split}
\end{equation}
Now we must define a set of projection operators
\begin{equation}\label{pplus}
	\mbox{p}_+^q\,:\,E_+\rar E_+^q\,\, (q=1,...,I),
\end{equation}
between the spaces defined in \eqref{E_0q} and \eqref{E_0}, which effectively separate out each $r^q$ term in the equations, and apply $\mbox{p}_+^{\tilde{k}}$ \eqref{pplus} to equations \eqref{slimkinE} for each $k\in\mc{E}$. This gives
\begin{equation}\label{vpsiprime}
	\begin{split}
		rv'_{k+1}=&-2(k+1)v_{k+1}-\frac{2}{r\mu}\left(m-\frac{P}{r}+\frac{r^3}{\ell^2}-\frac{r^3\eta}{\mu S^2}\right)v_{k+1}+\frac{k(k+1)}{r}\left(\frac{1}{\mu}-1\right)u_{k+1}\\
		&-\frac{2}{r^2\mu}\left(m-\frac{P}{r}+\frac{r^3}{\ell^2}-\frac{r^3\eta}{\mu S^2}\right)(k+1)u_{k+1}-\frac{r^2}{\mu S^2}\mbox{p}_+^{\tilde{k}}\hat{C}_k\\
		&-\frac{1}{r^{k+1}\mu}\slim_{s=0}^{N_1-2}\mbox{p}_+^{\tilde{k}}(f_{s+2})r^{s+1},\\
		rq'_k=&-2(k+1)q_k+\left(\frac{1}{\mu}-1\right)\frac{k(k+1)}{r}\psi_k-\frac{2}{r}\left(G+\frac{\zeta}{\mu^2 S^2}\right)\left(rq_k+k\psi_k\right)\\
		&+\frac{1}{r^{k}\mu}\slim_{s=1}^{N_2-2}\mbox{p}_+^{\tilde{k}}(g_{s+2})r^{s+1},
	\end{split}
\end{equation}
for all $k\in\mc{E}$. 

The main hurdle in rewriting this equation in a form to which Theorem \ref{Kthm3} may be applied is the final term in each of \eqref{vpsiprime}, as was the case for $\sun$ \cite{baxter_existence_2008, kunzle_analysis_1994}. As written here, it contains terms of much larger negative order than we want, i.e. terms of order $r^{-s}$ where $s>0$. Happily we may rewrite the final term using the following equalities. The first, concerning the magnetic gauge equations, is proven true in \cite{oliynyk_local_2002}:
\begin{equation}\label{frewrite}
	\frac{1}{r^{k+1}\mu}\slim_{s=0}^{N_1-2}\mbox{p}_+^{\tilde{k}}(f_{s+2})r^{s+1}=\frac{1}{\mu}\slim_{s=k}^{N_1-2}\mbox{p}_+^{\tilde{k}}(f_{s+2})r^{s-k}.
\end{equation}
The second is very similar and makes use of Theorem \ref{MyThrr0}:
\begin{equation}\label{grewrite}
	\frac{1}{r^{k}\mu}\slim_{s=1}^{N_2-2}\mbox{p}_+^{\tilde{k}}(g_{s+2})r^{s+1}=\frac{r}{\mu}\slim_{s=k}^{N_2-2}\mbox{p}_+^{\tilde{k}}(g_{s+2})r^{s-k}.
\end{equation}
The derivation of both \eqref{frewrite} and \eqref{grewrite} are very similar, so we shall describe how to derive them simultaneously by using the results from Section \ref{sec:sl2c}. Using Proposition \ref{Kprop4} and equation \eqref{fdef}, we may show that $f_k,g_k\in E_+\,\,\forall k$. From our definitions of the functions $u_{s+1}(r)$ and $\psi_s(r)$, we may see that $\chi_{s+1}u_{s+1},\chi_{s+1}\psi_s\in\bigoplus_{q=1}^{\tilde{s}}E_+^q$ for $0\leq s\leq \kappa_I$. Now we employ Theorem \ref{MyThrr0}, taking $Z_{1,0}=0$, $Z_{2,0}=\Omega_+$, $Z_{1,k}=\chi_{k+1}\psi_{k}$ and $Z_{2,k+1}=\chi_{k+1}u_{k+1}$ for $k\geq 0$, and it is clear that $f_{s+2},g_{s+2}\in\bigoplus_{q=1}^{\tilde{s}}E_+^q$. Thus,
\begin{equation}
	\mbox{p}_+^{\tilde{k}}(f_{s+2})=\mbox{p}_+^{\tilde{k}}(g_{s+2})=0\mbox{   if }s<k,\,\,\forall k\in\mc{E},
\end{equation}
because if $k\in\mc{E}$, then $k=\kappa_{\tilde{k}}$ and so if $s<k=\kappa_{\tilde{k}}$, then $\tilde{s}<\tilde{k}$, proving \eqref{frewrite} and \eqref{grewrite}.

Using \eqref{frewrite}, \eqref{grewrite} in \eqref{vpsiprime} and rearranging gives
\begin{equation}\label{rvprime}
	\begin{split}
		rv'_{k+1}=&-2(k+1)v_{k+1}-\frac{2}{r\mu}\left(m-\frac{P}{r}+\frac{r^3}{\ell^2}-\frac{r^3\eta}{\mu S^2}\right)v_{k+1}+\frac{k(k+1)}{r}\left(\frac{1}{\mu}-1\right)u_{k+1}\\
		&-\frac{2}{r^2\mu}\left(m-\frac{P}{r}+\frac{r^3}{\ell^2}-\frac{r^3\eta}{\mu S^2}\right)(k+1)u_{k+1}-\frac{r}{\mu}\slim_{s=k}^{N_1-2}\mbox{p}_+^{\tilde{k}}(f_{s+2})r^{s-k}\\
		&-\frac{r^2}{\mu S^2}\mbox{p}_+^{\tilde{k}}\hat{C}_k+\left(1-\frac{1}{\mu}\right)\mbox{p}_+^{\tilde{k}}(f_{k+2})-\mbox{p}_+^{\tilde{k}}(f_{k+2}),\\
		rq'_k=&-2(k+1)q_k+\left(\frac{1}{\mu}-1\right)\frac{k(k+1)}{r}\psi_k-\frac{2}{r}\left(G+\frac{\zeta}{\mu^2 S^2}\right)\left(rq_k+k\psi_k\right)\\
		&+\frac{r}{\mu}\slim_{s=k}^{N_2-2}\mbox{p}_+^{\tilde{k}}(g_{s+2})r^{s-k},
	\end{split}
\end{equation}
for all $k\in\mc{E}$. It is helpful to note that in this regime, $\frac{1}{\mu}-1=O(r^2)$, and that the boundary conditions \eqref{BCorigin} imply that $\hat{C}=O(r^2)$. Using the properties of $\HHIP{\,}{\,}$ and noting that $T_2(u_2)=2u_2$ and $T_2(\psi_1)=2\psi_1$, we can show that there exist analytic functions
\begin{equation}\label{PG1}
	\begin{array}{ccc}
		\hat{\eta}:E_0\times\R\rar\R, & \quad\quad & \hat{\zeta}:E_0\times E_+\times\R\rar\R,\\
		\hat{P}:E_+\times\mathbb{R}\rar\mathbb{R}, & \quad\quad & \hat{G}:E_+\times E_+\times\mathbb{R}\rar\mathbb{R},\\
	\end{array}
\end{equation}
with
\begin{equation}
	\begin{array}{ccc}
		\eta=-2\norm{\hat{a}_1}^2+r\hat{\eta}(\psi,q,r), & \quad\quad & \zeta=2r^2\norm{\psi_1}^2+r^3\hat{\zeta}(u,\psi,r),\\[5pt]
		P=r^4\norm{u_2}^2+r^5\hat{P}(u,r), & \quad\quad & G=2r^2\norm{u_2}^2+r^3\hat{G}(u,v,r),\\
	\end{array}
\end{equation}
where $u=\sum_{s\in\mc{E}}u_{s+1}$ and similarly for $v$; $\psi=\sum_{s\in\mc{E}}\psi_{s}$ and similarly for $q$; and $\norm{X}^2=\lnorm X|X\rnorm$. In writing $\eta$ we used the basis \eqref{defAhat} -- using \eqref{wandE} in \eqref{defEhat} we may see that $\psi_1$ and $\hat{a}_1$ are related by the linear transform
\begin{equation}
	\psi_1=[\hat{a}_1,\Omega_+].
\end{equation}
The important point here is that $\eta\sim O(1)$ near $r=0$, as suggested by the lower order terms of the expansion \eqref{prop0exp}. We also introduce a few more analytic maps for convenience:
\begin{equation}
	\begin{array}{ll}
		\mu^{-1}=1+r\hat{\mu}_A,\quad\quad & \mu^{-2}=1+r\hat{\mu}_B,\\[5pt]
		S^{-1}=1+r\hat{S}_A,\quad\quad & S^{-2}=1+r\hat{S}_B.
	\end{array}
\end{equation}
Now we rewrite the Einstein equations \eqref{EEs}. We introduce a new mass variable
\begin{equation}\label{curlyMdef}
	\mc{M}=\frac{1}{r^3}\left(m-r^3\left(\norm{u_2}^2+\frac{2}{3}\left(\norm{\psi_1}^2-\norm{\hat{a}_1}^2\right)\right)\right).
\end{equation}
We know that $\norm{u_2}$ and $\norm{\psi_1}$ (and hence $\norm{\hat{a}_1}$) are well defined since for all Lie groups, $\min\mc{E}=1$ and hence $\kappa_1=1$ always. Equations \eqref{EEs} then become
\begin{equation}\label{rMprime}
	\begin{split}
		r\mc{M}^\prime=&-3\mc{M}+r\left[\hat{P}(u,r)+\hat{G}(u,v,r)+\hat{\eta}(\psi,q,r)+\hat{\zeta}(u,\psi,r)-2\norm{\hat{a}_1}^2\hat{S}_B\right.\\
		&\left.+2\norm{\psi_1}^2\left(\hat{S}_B+\hat{\mu}_A\right)-2\lnorm u_2 | v_2 \rnorm+\frac{4}{3}\left(\lnorm \hat{a}_1 | \hat{a}_1^\prime \rnorm-\lnorm \psi_1 | q_1 \rnorm\right)\right]\\
		&+r^2\left[\left(\frac{1}{\ell^2}-2\mc{M}-2\norm{u_2}^2+\frac{4}{3}\left(\norm{\hat{a}_1}^2-\norm{\psi_1}^2\right)\right)\left(2\norm{u_2}^2+r\hat{G}\right)\right.\\
		&\left.+2\norm{\psi_1}^2\hat{\mu}_A\hat{S}_B+\hat{\zeta}\left(\hat{\mu}_A+\hat{S}_B\right)+\frac{2\norm{u_2}^2}{\ell^2}\right]+r^3\hat{\mu}_A\hat{S}_B\hat{\zeta},\\
		rS^\prime=&\,\;r^2S\left\{4\norm{u_2}^2+4\norm{\psi_1}^2+2r\left(\hat{\zeta}(u,\psi,r)+\hat{G}(u,v,r)+2\norm{\psi_1}^2\left(\hat{\mu}_B+\hat{S}_A\right)\right)\right.\\
		&\left.+r^2\left(2\norm{\psi_1}^2\hat{\mu}_B\hat{S}_A+\hat{\zeta}\left(\hat{\mu}_B+\hat{S}_A\right)\right)+r^3\hat{\mu}_B\hat{S}_A\hat{\zeta}\right\}.\\
	\end{split}
\end{equation}
We make one last variable change:
\begin{equation}\label{vhatdef}
	\hat{v}_{k+1}=v_{k+1}+\frac{1}{2(k+1)}\mbox{p}_+^{\tilde{k}}(f_{k+2}).
\end{equation}
To continue, we define $\bv{I}_\epsilon(0)$ as an open interval of size $|2\epsilon|$ on the real line about the point $0\in\R$:
\begin{equation}
	\bv{I}_\epsilon(0)\equiv(-\epsilon,\epsilon)
\end{equation}
where for our purposes, $\epsilon>0$ is small. We proceed by fixing vectors $X_1,X_2\in E_+$ and define $\hat{v}=\sum_{s\in\mc{E}}\hat{v}_{s+1}$. Then from (\ref{rvprime}, \ref{curlyMdef}, \ref{vhatdef}), we can show there exist neighbourhoods $\mc{N}_i$ of $X_i\in E_+$ with $i\in\{1,2\}$, some $\epsilon>0$, and a sequence of analytic maps
\begin{equation}
	\begin{split}
		\mc{Q}_k:\mc{N}_1\times E_+\times \bv{I}_\epsilon(0)\times \bv{I}_\epsilon(0)&\rar E_+^{\tilde{k}},\\
		\mc{G}_k:\mc{N}_2\times E_+\times \bv{I}_\epsilon(0)\times \bv{I}_\epsilon(0)&\rar E_+^{\tilde{k}},
	\end{split}
\end{equation}
for which
\begin{equation}\label{rhatvprime}
	\begin{split}
		rq'_k&=-2(k+1)q_k+r\mc{Q}_k(u,\psi,S,\mc{M},q),\\
		r\hat{v}'_{k+1}&=-2(k+1)\hat{v}_{k+1}+r\mc{G}_k(u,\hat{v},\mc{M},r),
	\end{split}
\end{equation}
for all $k\in\mc{E}$. Also, with (\ref{rMprime}, \ref{vhatdef}) and using $v_{s+1}=u'_{s+1}$ and $q_s=\psi'_s$, there exist analytic maps
\begin{equation}
	\begin{split}
		\mc{H}_k: & E_0\times E_+\times\R\rar E_+^{\tilde{k}},\\
		\mc{I}_k: & E_+\times E_+\times \R\rar E_+^{\tilde{k}},\\
		\mc{J}: & E_+\times E_+\times \mathbb{R}\times\mathbb{R}\rar\mathbb{R},\\
		\mc{K}: & E_+\times E_+\times \mathbb{R}\times\mathbb{R}\rar\mathbb{R},
	\end{split}
\end{equation}
for all $k\in\mc{E}$, such that
\begin{equation}\label{ruprime_etc}
	\begin{split}
		& r\psi'_k=r\mc{H}_k(u,S,\psi,q),\\
		& ru'_{k+1}=r\mc{I}_k(u,\hat{v}),\\
		& r\mc{M}^\prime=-3\mc{M}+r\mc{J}(u,\hat{v},\mc{M},r),\\
		& rS^\prime=r^2\mc{K}(u,\hat{v},S,r).\\
	\end{split}
\end{equation}
Now equations (\ref{rhatvprime}, \ref{ruprime_etc}) are in a form appropriate to Theorem \ref{Kthm3}. For fixed $X_1,X_2\in E_+$ there exists a unique solution
\begin{equation}
	\{\psi_k(r,Y_1),q_k(r,Y_1),u_{k+1}(r,Y_2),\hat{v}_{k+1}(r,Y_2),\mc{M}(r,Y_1,Y_2),S(r,Y_1,Y_2)\},
\end{equation}
analytic in a neighbourhood of $(r,Y_1,Y_2)=(0,X_1,X_2)$, satisfying
\begin{equation}\label{final0BCs}
	\begin{split}
		\psi_s(r,Y_1)&=rY_{1,s}+O(r^2),\\
		q_s(r,Y_1)&=O(r),\\
		u_{s+1}(r,Y_2)&=Y_{2,s}+O(r),\\
		\hat{v}_{s+1}(r,Y_2)&=O(r),\\
		\mc{M}(r,Y_1,Y_2)&=O(r),\\
		S(r,Y_1,Y_2)&=S_0+O(r^2),\\
	\end{split}
\end{equation}
for all $s\in\mc{E}$, where $Y_{i,s}=\mbox{p}_+^{\tilde{s}}(Y_i)$ for $i\in\{1,2\}$. From the definition of $\mc{M}$ \eqref{curlyMdef}, we can show that $m(r)=O(r^3)$. We note that $S_0$ is fixed by scaling such that $S_\infty=1$. Also, it is easy to see from (\ref{PG1}, \ref{vhatdef}, \ref{final0BCs}) that
\begin{equation}
	\eta=O(1),\quad\quad\zeta=O(r^2),\quad\quad P=O(r^4),\quad\quad G=O(r^2).
\end{equation}
From Lemma \ref{Klem2}, there exists an orthonormal basis $\{\bv{z}_j|j=1,...,\bsym{\Sigma}\}$ for $E_+$ consisting of the eigenvectors of $T_2$, i.e. $T_2(\bv{z}_j)=k_j(k_j+1)\bv{z}_j$. We introduce new variables in this basis:
\begin{equation}\label{usexp}
	\slim_{s\in\mc{E}}u_{s+1}(r)r^{s+1}=\slim_{j=1}^{\bsym{\Sigma}}\hat{u}_{j+1}(r)r^{k_j+1}\bv{z}_j,\quad \slim_{s\in\mc{E}}\psi_{s}(r)r^{s}=\slim_{j=1}^{\bsym{\Sigma}}\hat{\psi}_{j}(r)r^{k_j}\bv{z}_j.
\end{equation}
From Proposition \ref{Kprop1}, we know that $\bsym{\Sigma}=|\Sigma_w|$, so we can write $\Sigma_w=\{\alpha_j|j=1,...,\bsym{\Sigma}\}$; and from Proposition \ref{Kprop4}, we find that $\{\bv{e}_{\alpha_j}|j=1,...,\bsym{\Sigma}\}$ is also a basis for $E_+$. Therefore we can write
\begin{equation}\label{wvecdef}
	\bv{z}_j=\slim_{k=1}^{\bsym{\Sigma}} Q_{kj}\bv{e}_{\alpha_k}.
\end{equation}
With this definition of the matrix $Q_{ij}$, it is clear that the columns of $Q_{ij}$ are the eigenvectors of $T_2$. Now we expand $\Omega_+$, $W_+(r)$ and $\hat{E}_+(r)$ in the same basis:
\begin{equation}\label{omegabase}
	\Omega_+=\slim_{j=1}^{\bsym{\Sigma}}\omega_{j,0}\bv{e}_{\alpha_j},\quad W_+(r)=\slim_{j=1}^{\bsym{\Sigma}}\omega_j(r)\bv{e}_{\alpha_j},\quad \hat{E}_+(r)=\slim_{j=1}^{\bsym{\Sigma}}w^{\frac{1}{2}}_{j}\mc{E}_j(r)\bv{e}_{\alpha_j}.
\end{equation}
Then equations (\ref{K1}, \ref{usexp}, \ref{wvecdef}, \ref{omegabase}) imply that
\begin{equation}
	\omega_i(r)=\omega_{i,0}+\slim_{j=1}^{\bsym{\Sigma}} Q_{ij}\hat{u}_j(r)r^{k_j+1},\quad\mc{E}_i(r)=\slim_{j=1}^{\bsym{\Sigma}}Q_{ij}\hat{\psi}(r)r^{k_j}.
\end{equation}
for $i=1,...,\bsym{\Sigma}$ and with $\omega^2_{i,0}=w_i$. Finally, from \eqref{final0BCs} and \eqref{usexp} we find that
\begin{equation}
	\hat{\psi}_j(r,Y_1)=r\beta_j(Y_1)+O(r^2),\quad\hat{u}_j(r,Y_2)=\beta_j(Y_2)+O(r),\quad j=1,...,\bsym{\Sigma},
\end{equation}
with $\beta_j(Y_i)\equiv\lnorm\bv{z}_j|Y_i\rnorm$, $i\in\{1,2\}$. Therefore, we obtain the expansions \eqref{prop0exp}. $\Box$

\subsection{Local existence at the event horizon $r=r_h$}\label{sec:LocExRH}

\subsubsection{Boundary conditions at $r=r_h$}\label{ssec:bcsrh}

We use the notation $f_h\equiv f(r_h)$. For a regular non-extremal event horizon, we require $\mu_h=0$ and $\mu'_h$ finite and positive. This severely restricts the solution parameters and reduces the degrees of freedom of solutions, which makes boundary conditions easier to determine.

Transforming to a new variable $\rho=r-r_h$, we find that for regularity,
\begin{equation}\label{BCrh}
	\begin{split}
		\mu(\rho)&=\mu^\prime_h\rho+O(\rho^2),\\
		S(r)&=S_h+O(\rho),\\
		\omega_j(\rho)&=\omega_{j,h}+O(\rho),\\
		a_j(\rho)&=a^\prime_{j,h}\rho+O(\rho^2).
	\end{split}
\end{equation}
The constraint $\mu_h=0$ implies that
\begin{equation}
	\begin{split}
		m_h&=\frac{r_h}{2}+\frac{r^3_h}{2\ell^2},\qquad		\omega^\prime_{j,h}=-\frac{\mc{F}_h}{2\left(m_h-r_h^{-1}P_h+r_h^3\ell^{-2}-r_h^3\eta_h S_h^{-2}\right)},
	\end{split}
\end{equation}
with
\begin{equation}
	\mc{F}_h=\frac{1}{2}\slim_{i,j=1}^\mc{L}\omega_{i,h}C_{ij}(kw_j-\omega_{j,h}^2).
\end{equation}
We find that $\mu^\prime_h$  is given by
\begin{equation}\label{muph}
	\mu_h^\prime=\frac{k}{r_h}+\frac{3r_h}{\ell^2}-\frac{2P_h}{r_h^3}-\frac{2r_h\eta_h}{S_h^2}.
\end{equation}
The condition $\mu^\prime_h>0$ places an upper bound on $m^\prime_h$,
\begin{equation}\label{mhprime}
	m^\prime_h=\frac{P_h}{r_h^2}+\frac{r_h^2\eta_h}{S_h^2}<\frac{k}{2}+\frac{3r_h^2}{2\ell^2},
\end{equation}
with
\begin{equation}
	\begin{split}
		P_h&=\frac{1}{8}\slim_{i,j=1}^\mc{L}(kw_i-\omega_{i,h}^2)C_{ij}(kw_j-\omega_{j,h}^2),\qquad\eta_h=\disfrac{1}{2}\slim_{i,j=1}^\mc{L}a^\prime_{i,h}h_{ij}a^\prime_{j,h},
\end{split}
\end{equation}
and therefore with \eqref{mhprime} $m'_h$ also places a weak bound on the possible values of $\omega_{j,h}$ and $a^\prime_{j,h}$. We also notice that for $k=-1$, \eqref{mhprime} implies we have a minimum event horizon radius
\begin{equation}\label{minrh}
	r_h^2>\frac{\ell^2}{3}(2m_h'+1)>0,
\end{equation}
and a minimum value for $|\Lambda|$,

\begin{equation}
	|\Lambda|>\frac{1}{r_h^2}\left(1+2P_h+\frac{2r_h^2\eta_h}{S_h^2}\right).
\end{equation}

Fixing $r_h$ and $\ell$, and regarding $S_h$ as fixed by the requirement that the solution is asymptotically adS, the solution parameters are $\{\mc{E}_{j,h}^\prime,\omega_{j,h}\}$. Therefore, as at the origin, we have $2\mc{L}$ solution degrees of freedom at the event horizon. 

\subsubsection{Proof of local existence at $r=r_h$}

We begin by defining the new variable
\labeq{\rho=r-r_h}{rhodef}
so that for $r\rar r_h$ we are considering the limit $\rho\rar0$. 
We note that as in the asymptotically flat \cite{oliynyk_local_2002} and the purely magnetic adS cases \cite{baxter_general_2016}, we do not need the results of Section \ref{sec:sl2c} and use the notation $E_+$ out of convenience -- we could equally replace $E_+$ everywhere in the following with $\sum_{\alpha\in\Sigma_\lambda}\R\bv{e}_\alpha$, without using $E_+=\sum_{\alpha\in\Sigma_\lambda}\R\bv{e}_\alpha$. 
\begin{prop}\label{prop:lexrh}
	In a neighbourhood of the event horizon $r=r_h$ (i.e. $\rho=0$), with $r_h\neq0$, there exist regular black hole solutions to the field equations \eqref{EEs}, \eqref{YMEs}, analytic and unique with respect to their initial values, of the form
	\begin{equation}
		\begin{split}
			\mu(\rho)=&\mu^\prime_h\rho+O(\rho^2),\\
			S(\rho)=& S_h+O(\rho),\\
			\omega_j(\rho)=&\omega_{j,h}+O(\rho),\\
			\mc{E}_j(\rho)=& \mc{E}^\prime_{i,h}\rho+O(\rho^2)
		\end{split}
	\end{equation}
	with $\omega_{j,h}$, $\mc{E}^\prime_{j,h}$ arbitrary and $\mu^\prime_h$ given by \eqref{muph}.
\end{prop}

\textbf{Proof} We transform the field variables thus:
\begin{subequations}
	\begin{align}
		\bar{\lambda}(\rho)&=\frac{\mu(\rho)}{\rho}-\nu,\label{newmu}\\
		V_+(\rho)&=\frac{\mu W_+^\prime}{\rho},\label{Vplus}\\
		\Upsilon(\rho)&=\frac{A}{\rho},\\
		\Psi(\rho)&=\frac{r^2}{S}A^\prime,
	\end{align}
\end{subequations}
where $\nu$ is some constant we have yet to determine. Immediately we have
\begin{equation}\label{rhodw}
	\begin{array}{lcl}
		\rho\disfrac{dW_+}{d\rho}=\rho\left(\disfrac{V_+}{\bar{\lambda}+\nu}\right), & \quad & 
		\rho\disfrac{d\Upsilon}{d\rho}=-\Upsilon+\frac{S\Psi}{\rho^2};
	\end{array}
\end{equation}
and it is clear that there exist analytic maps $\hat{\mc{F}}:E_+\rar E_+$, $\hat{P}:E_+\rar\mathbb{R}$, with
\begin{equation}
	\hat{\mc{F}}(W_+)=\mc{F},\quad\quad\hat{P}(W_+)=P.
\end{equation}
We also notice that
\begin{equation}
	\eta=-\frac{S^2}{r^4}(\Psi^\prime,\Psi^\prime),\qquad\zeta=-\rho^2([\Upsilon,W_+],[\Upsilon,W_-]).
\end{equation}
So we define some more analytic maps, $\hat{\eta}:E_0\rar\R$, $\hat{\zeta}:E_0\times E_+\rar\R$, $\hat{G}:E_+\times \bv{I}_{|\nu|}(0)\rar\mathbb{R}$, $\hat{\mc{C}}:E_0\times E_+\rar E_+$ and $\hat{\mc{D}}:E_0\times E_+\rar E_+$ by
\begin{equation}
	\begin{split}
		\hat{\eta}(\Psi)&\equiv-\|\Psi^\prime\|^2,\\
		\hat{\zeta}(\Upsilon,W_+)&\equiv-([\Upsilon,W_+],[\Upsilon,W_-]),\\
		\hat{G}(X,a)&\equiv\frac{1}{2(a+\nu)^2}\|X\|^2,\\
		\hat{\mc{C}}(\Upsilon,W_+)&\equiv[\Upsilon,[\Upsilon,W_+]],\\
		\hat{\mc{D}}(\Upsilon,W_+)&\equiv[W_+,[\Upsilon,W_-]].
	\end{split}
\end{equation}
Then we can see that $G=\hat{G}(V_+,\bar{\lambda})$. Using these we can rewrite the EYM equations \eqref{YMEs} as
\begin{subequations}\label{dlamdv}
	\begin{align}
		\rho\frac{d\bar{\lambda}}{d\rho}=&\,-(\bar{\lambda}+\nu)+\frac{k}{r_h}+\frac{3r_h}{\ell^2}-\frac{2}{r_h^3}\hat{P}(W_+)+\frac{2\norm{\Psi}^2}{r_h^3}\\
		&+\rho\left[\frac{3}{\ell^2}+\frac{k}{\rho}\left(\frac{1}{\rho+r_h}-\frac{1}{r_h}\right)-\frac{2}{\rho}\left(\frac{1}{(\rho+r_h)^3}-\frac{1}{r_h^3}\right)\hat{P}(W_+)\right.\nonumber\\
		&\left.+\left(\frac{\bar{\lambda}+\nu}{\rho+r_h}\right)\left(1+2\hat{G}(V_+,\bar{\lambda})\right)+\frac{2\hat{\eta}(\Psi)}{\rho+r_h}+\frac{2\hat{\zeta}(\Upsilon,W_+)}{S^2\left(\bar{\lambda}+\nu\right)(\rho+r_h)}\right],\nonumber\\
		\rho\frac{dV_+}{d\rho}=&\,-V_+-\frac{1}{(\rho+r_h)^3}\hat{\mc{F}}(W_+)\nonumber\\
		&+\rho \left[-\frac{2V_+\hat{G}(V_+,\bar{\lambda})}{\rho+r_h}-\frac{2V_+\hat{\zeta}(\Upsilon,W_+)}{S^2\left(\bar{\lambda}+\nu\right)(\rho+r_h)}+\frac{\hat{\mc{C}}(\Upsilon,W_+)}{S^2\left(\bar{\lambda}+\nu\right)}\right],\\
		\rho\frac{dS}{d\rho}=&\,\rho\left[\frac{2S\hat{G}(V_+,\bar{\lambda})}{\rho+r_h}+\frac{2\hat{\zeta}(\Upsilon,W_+)}{S(\rho+r_h)(\bar{\lambda}+\nu)^2}\right],\\
		\rho\frac{d\Psi}{d\rho}=&\,\rho\left[\disfrac{\hat{\mc{D}}(\Upsilon,W_+)}{S\left(\bar{\lambda}+\nu\right)}\right].
	\end{align}
\end{subequations}
In order to cast the equations in the form necessary for Theorem \ref{Kthm3}, we introduce some final new variables:
\begin{subequations}\label{newvarrh}
	\begin{align}
	\tilde{\lambda}=&\bar{\lambda}+\nu-\frac{k}{r_h}+\frac{2}{r_h^3}\hat{P}(W_+)-\frac{3r_h}{\ell^2}+\frac{2}{r_h^3}\hat{\eta}(\Psi),\label{lamhat}\\
	\tilde{V}_+=& V_++\frac{1}{r_h^2}\hat{\mc{F}}(W_+),\\
	\tilde{\Upsilon}=&\Upsilon-\frac{S\Psi}{x^2}.
	\end{align}
\end{subequations}
We continue by defining an analytic map $\gamma:E_0\times E_+\times\mathbb{R}\rar\mathbb{R}$ with
\begin{equation}
	\gamma(X_1,X_2,a)=a-\nu+\frac{k}{r_h}-\frac{2}{r_h^3}\hat{P}(X_2)+\frac{3r_h}{\ell^2}-\frac{2}{r_h^3}\hat{\eta}(X_1).
\end{equation}
Fix vectors $Y_1\in E_0$, $Y_2\in E_+$, satisfying
\begin{equation}
	\|kr_h^{-1}-2r_h^{-3}\hat{P}(Y_2)+3r_h\ell^{-2}-2r_h^{-3}\hat{\eta}(Y_1)\|>0.
\end{equation}
Then if we set
\begin{equation}\label{nudefrh}
	\nu=\frac{k}{r_h}+\frac{3r_h}{\ell^2}-\frac{2}{r_h^3}\hat{P}(Y_2)-\frac{2}{r_h^3}\hat{\eta}(Y_1),
\end{equation}
it is obvious that $\gamma(Y_1,Y_2,0)=0$. Therefore, define an open neighbourhood $D$ of $(Y_1,Y_2,0)\in E_0\times E_+\times\mathbb{R}$ by
\begin{equation}
	D=\{(X_1,X_2,a)\,|\,\|\gamma(X_1,X_2,a)\|<\|\nu\|\}.
\end{equation}
Then from \eqref{rhodw}, \eqref{dlamdv}, \eqref{newvarrh} we can show there must exist some $\epsilon>0$ and analytic maps
\begin{equation}\label{rhfin}
	\begin{array}{rcl}
		\mc{G}: E_+\times D\rar\mathbb{R} & \mbox{with} & \rho\disfrac{dW_+}{d\rho}=\rho\,\mc{G}(\tilde{V}_+,W_+,\tilde{\lambda}),\\[10pt]
		\mc{H}: E_0\times E_+\times D\times \bv{I}_\epsilon(0)\rar\mathbb{R} & \mbox{with} & \rho\disfrac{d\tilde{V_+}}{d\rho}=-\tilde{V}_++\rho\mc{H}(\tilde{\Upsilon},\tilde{V}_+,W_+,\tilde{\lambda},\rho),\\[10pt]
		\mc{K}: E_+\times D\times \bv{I}_\epsilon(0)\rar\mathbb{R} & \mbox{with} & \rho\disfrac{d\tilde{\lambda}}{d\rho}=-\tilde{\lambda}+\rho\mc{K}(\tilde{V}_+,W_+,\tilde{\lambda},\rho),\\[10pt]
		\mc{L}: E_0\times E_+\times\R\times\bv{I}_\epsilon(0)\rar\R & \mbox{with} & \rho\disfrac{dS}{d\rho}= \rho\mc{L}(\tilde{\Upsilon},\tilde{V}_+,S,\rho),\\[10pt]
		\mc{M}: E_0\times\R\times\bv{I}_\epsilon(0)\rar\R & \mbox{with} & \rho\disfrac{d\tilde{\Upsilon}}{d\rho}=-\tilde{\Upsilon}+\rho\mc{M}(\Psi,S,\rho),\\[10pt]
		\mc{N}: E_0\times E_+\times D\times\R\times\bv{I}_\epsilon(0)\rar\R & \mbox{with} &  \rho\disfrac{d\Psi}{d\rho}=\rho\mc{N}(\tilde{\Upsilon},W_+,\tilde{\lambda},S,\rho).
	\end{array}
\end{equation}
It can be seen that \eqref{rhfin} are in the form applicable to Theorem \ref{Kthm3}. Hence there is a unique solution
\begin{equation}\nonumber
	\{\tilde{\Upsilon}(\rho,U_1,U_2),\Psi(\rho,U_1,U_2),W_+(\rho,U_1,U_2),\tilde{V}_+(\rho,U_1,U_2),\tilde{\lambda}(\rho,U_1,U_2), S(\rho,U_1,U_2)\},
\end{equation}
analytic in a neighbourhood of $(\rho,U_1,U_2)=(0,Y_1,Y_2)$, which satisfies
\begin{equation}
	\begin{array}{lcl}
		\tilde{\Upsilon}(\rho,U_1,U_2)=Y_1\rho+O(\rho^2), & \,\, & \Psi(\rho,U_1,U_2)=O(\rho).\\
		W_+(\rho,U_1,U_2)=Y_2+O(\rho),\label{finw} 		  & \,\, & \tilde{V}_+(\rho,U_1,U_2)=O(\rho),\\
		\tilde{\lambda}(\rho,U_1,U_2)=O(\rho), 			  & \,\, & S(\rho,U_1,U_2)=S_h+O(\rho).
	\end{array}
\end{equation}
To gain a more explicit solution, we expand $Y_1$, $Y_2$, $A$, $W_+$ in the basis $\{\bv{h}_i,\bv{e}_\alpha | i\in\{1,...,\bsym{\Sigma}\},\,\alpha\in \Sigma_w\}$, as follows:
\begin{equation}\label{rhbase}
	\begin{array}{ccc}
		Y_1=\slim_{i=1}^\mc{L}a^\prime_{i,h}\bv{h}_i, & \quad\quad & \,\,A(\rho)=\slim_{i=1}^\mc{L}a_{i}(\rho)\bv{h}_i,\\[10pt]
		Y_2=\!\!\!\slim_{\alpha\in \Sigma_w}\!\!\!\omega_{\alpha,h}\bv{e}_\alpha, & \quad\quad & W_+(\rho)=\!\!\!\slim_{\alpha\in \Sigma_w}\!\!\!\omega_{\alpha}(\rho)\bv{e}_\alpha.
	\end{array}
\end{equation}
Noting \eqref{finw}, this yields
\begin{equation}
	a_i(\rho,Y_1)=Y_1\rho+O(\rho^2),\qquad \omega_\alpha(\rho,Y_2)=Y_2+O(\rho),
\end{equation}
for all $\alpha\in \Sigma,i\in\{1,...,\mc{L}\}$; or using the basis \eqref{Ebar}, we can express the electric equation as
\begin{equation}
	\mc{E}_i(\rho,Y_1)=\slim_{\alpha\in\Sigma_w}[Y_1,\bv{e}_{\alpha}]\rho+O(\rho^2)
\end{equation}
Finally, it is easy to show from \eqref{newmu}, \eqref{lamhat}, \eqref{nudefrh} that
\begin{equation}
	\mu(\rho,Y_1,Y_2)=\nu\rho+O(\rho^2),\,\,\,\mbox{where}\,\,\mu_h=0,\,\mu'_h=\nu.\,\Box
\end{equation}

\subsection{Local existence as $r\rar\infty$}\label{sec:LocExInf}

\subsubsection{Boundary conditions as $r\rar\infty$}

We assume power series for all field variables of the form $f(r)=f_\infty+\sum_{i=1}^\infty f_ir^{-i}$ and for clarity use the base \eqref{basedef} and \eqref{Ebar}. Examining \eqref{EEs} and \eqref{YMEs}, we find that the expansions near infinity must be
\begin{equation}\label{BCinf}
	\begin{split}
		m(r)&=m_\infty+m_1r^{-1}+O(r^{-2}),\\
		S(r)&=S_\infty+S_4r^{-4}+O(r^{-5}),\\
		\omega_j(r)&=\omega_{j,\infty}+\omega_{j,1}r^{-1}+\omega_{j,2}r^{-2}+O(r^{-3}),\\
		\mc{E}_j(r)&=\mc{E}_{j,\infty}+\mc{E}_{j,1}r^{-1}+\mc{E}_{j,2}r^{-2}+O(r^{-3}).\\
	\end{split}
\end{equation}
The asymptotic power series expansions are as expected: no constraints are placed on $S_\infty$, $m_\infty$, so we set $m_\infty=M$ (the constant Arnowitt-Deser-Misner (ADM) mass) and rescale $S$ to $S_\infty=1$ to agree with the asymptotic limit for SadS (or pure adS space if $M=0$). There are also no constraints placed on $\omega_{j,\infty}$, $\omega_{j,1}$, $\mc{E}_{j,\infty}$ or $\mc{E}_{j,1}$ -- this is in accord with the discussion in Section \ref{sec:asymregreq}, and will be investigated further in Section \ref{sec:asym}. Furthermore we find that all higher order terms we calculate in the expansion are determined by lower order terms. The lowest order terms are
\begin{equation}
	\begin{split}
		m_1=&-\frac{1}{\ell^2}\slim_{j=1}^{\mc{L}}\frac{\omega_{j,1}^2}{|\alpha_j|^2}-\frac{1}{8}\slim_{j,l=1}^{\mc{L}}(kw_j-\omega_{j,\infty}^2)h_{jk}(kw_l-\omega_{l,\infty}^2)\\
		&-\slim_{i,j=1}^\mc{L}\frac{\mc{E}_{i,1}(h^{-1})_{ij}\mc{E}_{j,1}}{|\alpha_i|^2}-\ell^2\slim_{i=1}^\mc{L}\frac{\omega^2_{i,\infty}\mc{E}^2_{i,\infty}}{|\alpha_i|^2},\\
		S_4=&-\frac{\ell^4}{2}\slim_{i=1}^\mc{L}\frac{\omega^2_{i,\infty}\mc{E}^2_{i,\infty}}{|\alpha_i|^2}-\frac{1}{2}\slim_{i=1}^{\mc{L}}\frac{\omega_{i,1}^2}{|\alpha_i|^2},\\
		\omega_{j,2}=&\,\ell^2\omega_{j,\infty}\slim_{l=1}^{\mc{L}}C_{jl}(kw_l-\omega_{l,\infty}^2).\\
	\end{split}
\end{equation}
So our solution parameters asymptotically are $\{M,\omega_{j,\infty},\omega_{j,1},\mc{E}_{j,\infty},\mc{E}_{j,1}\}$ and we have therefore $4\mc{L}+1$ degrees of freedom in total.

\subsubsection{Proof of local existence as $r\rar\infty$}

In Section \ref{sec:asymregreq}, we thus confirmed there are no constraints on the asymptotic boundary for the gauge fields as there was for asymptotically flat solutions: this was the case in $\sun$ \cite{baxter_general_2016,baxter_existence_2016} and is to be expected. (We will come back to this point in Section \ref{sec:asym}.) Hence, as in Section \ref{sec:LocExRH}, we do not need any of the results of Section \ref{sec:sl2c} here, but we still use the notation $E_+$ for convenience. To deal sensibly with the limit $r\rar\infty$ we transform to the variable 
\begin{equation}\label{zr}
	z=r^{-1},
\end{equation}
whence we are now dealing with the limit $z\rar0$. 
\begin{prop}\label{prop:lexinf}
	There exist regular solutions of the field equations in some neighbourhood of $z=0$, analytic and unique with respect to their initial values, of the form
	\begin{equation}\label{propinfexp}
		\begin{split}
			m(z)&=M+O(z),\\
			S(z)&=1+O(z^4),\\
			\omega_i(z)&=\omega_{i,\infty}+\omega_{i,1}z+O(z^2),\\
			\mc{E}_i(z)&=\mc{E}_{i,\infty}+\mc{E}_{i,1}z+O(z^2),
		\end{split}
	\end{equation}
	where $\omega_{i,\infty}$, $\mc{E}_{i,\infty}$, $\omega_{i,1}$ and $\mc{E}_{i,1}$ are arbitrary, and in order to agree with the asymptotic limit of adS space, we have let $m_\infty=M$, the ADM mass of the solution, and $S_\infty=1$.
\end{prop}
\textbf{Proof} As well as \eqref{zr}, we introduce also the following new variables:
\begin{equation}\label{infvars}
	\begin{split}
		\lambda(z)=2m(r),\quad\quad v_0(z)=r^2A^\prime(r),\quad\quad v_+(z)=r^2W^\prime_+(r).
	\end{split}
\end{equation}
We immediately find that
\begin{equation}
	z\frac{dW_+}{dz}=-zv_+,\quad\quad z\frac{dA}{dz}=-zv_0
\end{equation}
and it is clear that there exist analytic maps:
\begin{equation}
	\begin{array}{rcl}
		\hat{\mc{F}}:E_+\rar E_+ 			& \mbox{with} & \hat{\mc{F}}(W_+)=\mc{F},\\
		\hat{\mc{Z}}:E_0\times E_+\rar E_+ 	& \mbox{with} & \hat{\mc{Z}}(A,W_+)=\mc{Z},\\
		\hat{\zeta}:E_0\times E_+\rar\R 	& \mbox{with} & \hat{\zeta}(A,W_+)=\zeta, \\
		\hat{P}:E_+\rar\mathbb{R} 			& \mbox{with} & \hat{P}(W_+)=P, \\
		\hat{\mc{Y}}:E_0\times E_+\rar E_+ 	& \mbox{with} & \hat{\mc{Y}}(A,W_+)=[A,[A,W_+]]. \\ 
	\end{array}
\end{equation}
Also we find that in this limit,
\begin{equation}\label{GLocEx}
	G=\frac{z^4}{2}(v_+,v_-),\quad\quad \hat{\zeta}\sim O(1),\quad\quad \hat{\mc{Y}}\sim O(1)\quad\quad \mu\sim k+\frac{1}{z^2\ell^2},
\end{equation}
which in particular implies that
\begin{equation}\label{mulim}
	\frac{1}{\mu z^2\ell^2}-1=O(z^2),\quad\mbox{ and }\quad \frac{1}{\mu z}=O(z).
\end{equation}
Then using \eqref{GLocEx} and \eqref{mulim}, we may see that
\begin{equation}\label{infeqs}
	\begin{split}
		z\frac{dS}{dz}=& z^4\left(\norm{v_+}^2S+\frac{2\hat{\zeta}(A,W_+)}{S^2}\right),\\
		z\frac{d\lambda}{dz}=&-2z\left(-\frac{\norm{v_0}^2}{s^2}+\frac{\hat{\zeta}(A,W_+)}{\mu zS^2}+\hat{P}(W_+)+\left(kz^2-\lambda z^3+\frac{1}{\ell^2}\right)\frac{\norm{v_+}^2}{2}\right),\\
		z\frac{dv_+}{dz}=&\,2v_+\left(\frac{1}{\mu z^2\ell^2}-1\right)+\frac{1}{\mu z}\left\{\hat{\mc{F}}(W_+)+v_+\left(\lambda z^2-2\hat{P}(W_+)z^3+\frac{2z^3}{S^2}\norm{v_0}^2\right)\right.\\
		&\left.+\frac{1}{\mu S^2z^2}\hat{\mc{Y}}(A,W_+)\right\},\\
		z\frac{dv_0}{dz}=& v_0\left(-z^4\norm{v_+}^2-\frac{\hat{\zeta}(A,W_+)}{2\mu^2 S^2}\right)-\frac{1}{\mu z}\hat{\mc{Y}}(A,W_+).
	\end{split}
\end{equation}
Then we fix four vectors $X_1,C_1\in E_0$, $X_2,C_2\in E_+$. From \eqref{infvars} -- \eqref{infeqs} it is clear that there exists an $\epsilon>0$ and analytic maps
\begin{equation}
	\begin{array}{rcl}
		\mc{G}_\infty\,:\,E_+\rar\R & \mbox{with} & z\disfrac{dW_+}{dz}=z\mc{G}_\infty(v_+),\\[10pt]
		\mc{H}_\infty\,:\,E_0\rar\R & \mbox{with} & z\disfrac{dA}{dz}=z\mc{H}_\infty(v_0),\\[10pt]
		\mc{I}_\infty\,:\,E_+\times\R\rar\R & \mbox{with} & z\disfrac{dS}{dz}=z^4\mc{I}_\infty(A,W_+,v_+,S),\\[10pt]
		\mc{J}_\infty\,:\,(E_0)^2\times (E_+)^2\times\R\times\bv{I}_\epsilon(0)\rar\R & \mbox{with} & z\disfrac{d\lambda}{dz}=z\mc{J}_\infty(A,W_+,v_0,v_+,\lambda,z),\\[10pt]
		\mc{K}_\infty\,:\,(E_0)^2\times (E_+)^2\times\R\times\bv{I}_\epsilon(0)\rar\R & \mbox{with} & z\disfrac{dv_+}{dz}=z\mc{K}_\infty(A,W_+,v_0,v_+,\lambda,z),\\[10pt]
		\mc{M}_\infty\,:\,(E_0)^2\times (E_+)^2\times\R\times\bv{I}_\epsilon(0)\rar\R & \mbox{with} & z\disfrac{dv_0}{dz}=z\mc{M}_\infty(A,W_+,v_0,v_+,\lambda,z)
	\end{array}
\end{equation}
%
%
%
(where we abbreviate $E_0\times E_0$ to $(E_0)^2$, and similar for $E_+$). Finally, Theorem \ref{Kthm3} says that these equations possess a unique solution analytic in some neighbourhood of $(z,Y_i,Z_i)=(0,X_i,C_i)$ (with $i\in\{1,2\}$) with behaviour
\begin{equation}\label{infexp}
	\begin{split}
		S(z,Y_i,Z_i)&=S_\infty+O(z^4),\\
		\lambda(z,Y_i,Z_i)&=\lambda_\infty+O(z),\\
		A(z,Y_i,Z_i)&=X_1+O(z),\\
		W_+(z,Y_i,Z_i)&=X_2+O(z),\\
		v_0(z,Y_i,Z_i)&=C_1+O(z),\\
		v_+(z,Y_i,Z_i)&=C_2+O(z),
	\end{split}
\end{equation}
and using the definitions \eqref{infvars},
\begin{equation}
	\begin{split}
		a_i(z,X_1,X_2,C_1,C_2)=X_1+C_1z+O(z^2),\\
		\omega_{\alpha}(z,X_1,X_2,C_1,C_2)=X_2+C_2z+O(z^2),
	\end{split}
\end{equation}
for all $\alpha\in \Sigma_w, i\in\{i,...,\mc{L}\}$. We expand the relevant vectors in \eqref{infexp} explicitly in the bases
\begin{equation}\label{182}
	\begin{array}{lll}
		X_1=\slim_{i=1}^\mc{L}a_{i,\infty}\bv{h}_i, & \quad C_1=\slim_{i=1}^\mc{L}a_{i,1}\bv{h}_i, & \quad \,\,\,\,\,\,A=\slim_{i=1}^\mc{L}a_i(z)\bv{h}_i,\quad\quad\\[15pt]
		X_2=\!\!\!\slim_{\alpha\in \Sigma_w}\!\!\!\omega_{\alpha,\infty}\bv{e}_\alpha, & \quad C_2=\!\!\!\slim_{\alpha\in \Sigma_w}\!\!\!\omega_{\alpha,1}\bv{e}_\alpha, & \quad W_+=\!\!\!\slim_{\alpha\in \Sigma_w}\!\!\!\omega_{\alpha}(z)\bv{e}_\alpha, 
	\end{array}
\end{equation}
to gain the familiar expansions
\begin{equation}
	\begin{split}
		a_i(z)&=a_{i,\infty}+a_{i,1}z+O(z^2),\\
		\omega_i(z)&=\omega_{i,\infty}+\omega_{i,1}z+O(z^2)
	\end{split}
\end{equation}
for all $i\in\{i,...,\mc{L}\}$. Once again, we may rewrite the electric gauge function in our basis \eqref{Ebar} as
\begin{equation}
	\mc{E}_i(z)=\mc{E}_{i,\infty}+\mc{E}_{i,1}z+O(z^2)
\end{equation}
where $\mc{E}_{i,\infty}\equiv\sum_{\alpha\in\Sigma_w}[X_1,\bv{e}_\alpha]$ and $\mc{E}_{i,1}\equiv\sum_{\alpha\in\Sigma_w}[C_1,\bv{e}_\alpha]$. Finally, we set $m_\infty=M$, $S_\infty=1$ for the asymptotically adS limit, and therefore we end up with the expansions \eqref{propinfexp}, having $4\mc{L}+1$ degrees of freedom. $\Box$

\section{Asymptotic behaviour of the field equations}\label{sec:asym}

As we saw in Section \ref{sec:asymregreq}, and further confirmed in Section \ref{sec:LocExInf}, the asymptotic boundary conditions \eqref{BCinf} imply that any regular solutions in the limit $r\rar\infty$ will have gauge functions which are characterised entirely by $\omega_{j,\infty}$, $\omega_{j,1}\equiv\omega^\prime_{j,\infty}$, $\mc{E}_{j,\infty}$, and $\mc{E}_{j,1}\equiv\mc{E}^\prime_{j,\infty}$, with all higher order terms in the expansions determined by these parameters. The reason for examining this is that the $\Lambda=0$ case is not so simple: there, the asymptotic values of the gauge field must approach a particular discrete set of values, and the higher order terms are intricately interdependent. Therefore we briefly digress to demonstrate the difference for $\Lambda<0$, which is highly similar to what we found in simpler cases \cite{baxter_existence_2015,baxter_existence_2016,baxter_existence_2008}.

So what we wish to do here is take the asymptotic limit of the field equations by transforming the system into autonomous form, and examining the phase plane of the system. The form of the parameter to which we must transform dictates the asymptotic behaviour of the field variables, and this gives us an infinitely richer solution space.
%

Firstly, we note that as $r\rar\infty$, $\mu\sim\frac{r^2}{\ell^2}$. Noting also \eqref{BCinf}, the Yang-Mills field equations \eqref{YMEs} become asymptotically,
\begin{equation}\label{YMasym}
	\begin{array}{ccc}
		r^2\left(\disfrac{r^2}{\ell^2}\omega^\prime_i\right)^\prime=-\mc{F}_i-\disfrac{\ell^2}{4}\omega_i\mc{E}_i^2, & \qquad & \disfrac{r^2}{\ell^2}\left(r^2\mc{E}^\prime_i\right)^\prime=\mc{Z}_i.
	\end{array}
\end{equation}
Using the parameter $\tau=\ell r^{-1}$, these are equivalent to
\begin{equation}
	\begin{array}{ccc}
		\disfrac{d^2\omega_i}{d\tau^2}=-\disfrac{1}{2}\slim_{j=1}^{\mc{L}}\omega_iC_{ij}(kw_j-\omega_{j}^2)-\frac{\ell^2}{4}\omega_i\mc{E}_i^2, & \qquad & \disfrac{d^2\mc{E}_i}{d\tau^2}=\slim_{j=1}^{\mc{L}}C_{ij}\omega_j^2\mc{E}_j.
	\end{array}
\end{equation}
It is clear that the critical points of this autonomous system, pairs $(\omega^*_i,\mc{E}_i^*)$, satisfy 
\begin{equation}
	\omega^*_i\left(\slim_{j=1}^{\mc{L}}C_{ij}(kw_j-\omega_j^{*2})+\frac{\ell^2}{2}\mc{E}_i^{*2}\right)=0,\qquad
	\slim_{j=1}^{\mc{L}}C_{ij}\omega_j^{*2}\mc{E}^*_j=0.
\end{equation}
This gives us only two sets of critical points in the 4D phase plane $\left(\omega_i,\mc{E}_i,\frac{d\omega_i}{d\tau},\frac{d\mc{E}_i}{d\tau}\right)$: the point $(0,\mc{E}_i^*,0,0)$ where $\mc{E}^*_i$ is arbitrary, and $(\pm \sqrt{kw_i},0,0,0)$, $\forall i$ ($i\in\{1,...,\mc{L}\}$), though it is obvious that for $k=-1$ the latter point does not exist, and for $k=0$ it coincides with the former if $\mc{E}_i^*=0$. 
However, we point out that we have used a parameter $\tau=\ell/r$ which compactifies our range of integration from $(r_h,\infty)$ to $(0,\ell\tau_h^{-1})$. Thus as we integrate out further and further, solution trajectories will in general not end at the critical points of the phase plane, but at some other more arbitrary value. In the asymptotically flat purely magnetic case, the parameter used was proportional to $\ln r$, so the integration domain is $(\ln(r_h),\infty)$ and thence every solution must end at a critical point.

This is why the analysis of the asymptotic boundary conditions \eqref{BCinf} for $\Lambda<0$ imply no constraints on the asymptotic values of $\omega_i(r)$ or $\mc{E}_i(r)$. In the purely magnetic spherical \cite{baxter_general_2016} and topological dyonic $\sun$ \cite{baxter_existence_2016} cases, it was this that was responsible for the existence of plentiful global solutions, only local existence could be established for $\Lambda=0$ \cite{oliynyk_local_2002}. To summarise, we have shown that as long as we can integrate a solution arbitrarily far into the asymptotic regime, it remains regular as $r\rar\infty$, reaching arbitrary asymptotic boundary values. We return to this point in Section \ref{sec:GloExArg}.

\section{Global existence proofs}\label{sec:GloReg}

Here we prove some results concerning the global behaviour of the solutions, which will culminate in the main results of this work, which is the proof of the global existence of non-trivial solutions to the field equations \eqref{EEs}, \eqref{YMEs}: firstly in some neighbourhood of the known trivial solutions from Section \ref{sec:embed} (Theorem \ref{prop:gloex}), and then in the limit $|\Lambda|\rar\infty$ (Theorem \ref{thr:l0}). First though, we prove a couple of necessary results concerning global behaviour.

\subsection{Proof that $\mc{E}_i(r)$ is monotonic $\forall i$}\label{sec:MonE}

Here we prove that the functions $\mc{E}_i$ are monotonically increasing in $r$ for all $i$. This proof is very similar to an analogous proof in \cite{baxter_existence_2016}, but we will give the main points.

We can write the electric gauge equation \eqref{YME} as
\begin{equation}\label{EGFPerfDeriv}
	\mu S\left(\frac{r^2\mc{E}_i^\prime}{S}\right)^\prime=\slim_{j=1}^\mc{L}C_{ij}\omega_j^2\mc{E}_j.
\end{equation}
Now $C_{ij}$ is a Cartan matrix, and so it has full rank and therefore $\mc{L}$ linearly independent eigenvectors, which we shall call $v_i=\{v_1,...,v_{\mc{L}}\}$. Furthermore, the eigenvalues are $\kappa_j(\kappa_j+1)$ for a series of integers $\kappa_j>0$ which depend on the Lie algebra $\mk{g}$ in question (See Table 1). Multiplying \eqref{EGFPerfDeriv} through by $v_i$, summing over $i$, and noticing that the eigenvectors are linearly independent, we find the system decouples into the $\mc{L}$ equations
%
%
%
\begin{equation}
	\left(\frac{r^2\mc{E}_i^\prime}{S}\right)^\prime=\kappa_i(\kappa_i+1)\frac{\omega_i^2\mc{E}_i}{\mu S},
\end{equation}
and integrating,
\begin{equation}
	\left[\frac{r^2\mc{E}^\prime_i}{S}\right]^{r_1}_{r_0}=\kappa_i(\kappa_i+1)\int\limits_{r_0}^{r_1}\frac{\omega_i^2\mc{E}_i}{\mu S}dr.
\end{equation}

We know that $\mu$, $S$, $r^2$, $\omega_i^2$ and $\kappa_i(\kappa_i+1)$ are non-negative, so the integrand is non-negative, and hence so is the integral. Therefore $\mc{E}_i(r)$ and $\mc{E}_i^\prime(r)$ have the same sign. Coupled to the fact that $\mc{E}_i(r)=0$ at the event horizon or origin, it is clear that each $\mc{E}_i(r)$ is always positive (negative) and monotonically increasing (decreasing) for all $i,\,r$. A corollary of this is that also, each $\mc{E}_i(r)$ must be non-zero for all $r>r_h$ (or $r>0$ for solitons).
%
%

\subsection{Global regularity of solutions}\label{sec:GloRegArg}

We prove here that any solution may be integrated out from the boundary $r=r_0$, where $r_0=r_h$ for black holes and $r_0=0$ for solitons, and will remain regular for $r$ arbitrarily large. This is conditional upon our metric function $\mu(r)$ being positive for all $r>r_0$. Thus we have:

\begin{prop}\label{prop:gloreg}
	If $\mu(r)>0\,\,\,\forall r>r_h$ for black holes (or $\forall r>0$ for solitons), then all field variables may be integrated out from the boundary conditions at the event horizon (or the origin) into the asymptotic regime, and will remain regular throughout.
\end{prop}
\textbf{Proof} Define $\mc{Q}\equiv[r_0,r_1)$ and $\bar{\mc{Q}}\equiv[r_0,r_1]$. The results of Section \ref{sec:LocEx} show that the field variables are regular at $r=r_0$. Our aim is therefore to use the fact that all field variables are regular on $\mc{Q}$, i.e. in a neighbourhood of $r=r_0$, and then show using the field equations that as long as the metric function $\mu(r)>0\,\forall r\in(r_0,\infty)$, then they will remain regular on $\bar{\mc{Q}}$ also, i.e. at $r=r_1$; and thus we can integrate the field equations out arbitrarily far and the field variables will remain regular. We note that this proof is completely independent of which basis we use, and so does not depend on our model being a regular model in the sense of \cite{brodbeck_generalized_1993,oliynyk_local_2002}.

First we recall that $\eta,\zeta,G,P\geq0$, so that $m'(r)\geq0\,\,\forall r$ and thus $m(r)$ is monotonically increasing, as expected. The same applies to $(\ln |S(r)|)^\prime$, showing that $\ln |S(r)|$ and hence $S(r)$ is monotonically increasing too. This means that (if the limits exist),
\begin{equation}
	m_{\max}\equiv\sup\{m(r)\,|\,r\in\bar{\mc{Q}}\}=m(r_1),\qquad S_{\max}\equiv\sup\{S(r)\,|\,r\in\bar{\mc{Q}}\}=S(r_1).
\end{equation}
The condition $\mu(r)\geq0\,\forall r\in[r_0,\infty)$ gives us our starting point, since with \eqref{BCrh} this implies that
\begin{equation}\label{mmax}
	m_{\max}\leq \frac{kr_1}{2}+\frac{r_1^3}{2\ell^2},
\end{equation}
giving us an absolute upper bound to work with -- for $k=-1$, the minimum event horizon radius \eqref{minrh} means that the right-hand side of \eqref{mmax} is non-negative. Therefore, both $m(r)$ and $\mu(r)$ are bounded on $\bar{\mc{Q}}$. This means that $m_{\max}$ exists, and for later use we define $\mu_{\min}\equiv\inf\{\mu(r)\,|\,r\in\bar{\mc{Q}}\}$.

Next we examine \eqref{EEm}. It is clear that
\begin{equation}
	2m^\prime(r)\geq 2\mu G+\frac{2\zeta}{\mu S^2},
\end{equation}
and integrating, we can show that
\begin{equation}
	\frac{2[m(r_1)-m(r_0)]}{\mu_{\min}}\geq \int\limits_{r_0}^{r_1}\left(2G+\frac{2\zeta}{\mu^2 S^2}\right)dr,
\end{equation}
which implies that $\ln|S|$ and hence $S$ is bounded on $\bar{\mc{Q}}$, so that $S_{max}$ also exists. It also implies that $G$ is bounded on $\bar{\mc{Q}}$, and since
\begin{equation}
	2G=\norm{W_+^\prime}^2
\end{equation}
then again by integrating and using the Cauchy-Schwartz inequality, we obtain
\begin{equation}\label{intG}
	\int\limits^{r_1}_{r_0}2Gdr=\int\limits^{r_1}_{r_0}\norm{W_+^\prime}^2dr\geq\left(\int\limits^{r_1}_{r_0}\norm{W_+^\prime}dr\right)^{\!\!2}=\left(\norm{W_+}\Big|_{r=r_1}-\norm{W_+}\Big|_{r=r_0}\right)^{\!\!2}.
\end{equation}
The left-hand side of \eqref{intG} is bounded and the right hand side is a sum of positive terms, and so $\norm{W_+}$ and hence $W_+$, $\mc{F}$ and $P$ are all bounded on $\bar{\mc{Q}}$.

In the same fashion,
\begin{equation}
	m^\prime(r)\geq\frac{r^2\eta}{S^2},
\end{equation}
so that
\begin{equation}\label{AA}
	\frac{S^2_{max}[m(r_1)-m(r_0)]}{r_0^2}\geq-\left(\norm{A}_{r=r_1}-\norm{A}_{r=r_0}\right)^2,
\end{equation}
(where we recall that the right-hand side of \eqref{AA} is \ul{positive} because $A$ is purely imaginary) so that it is obvious that $A$ is bounded on $\bar{\mc{Q}}$.

Finally, we take the gauge equations. We may rewrite \eqref{CommEqns} as
\begin{equation}
	\begin{split}
		\left(\mu SW^\prime_+\right)^\prime=&-\frac{S\mc{F}}{r^2}+\frac{1}{\mu S^2}[A,[A,W_+]],\\
		\left(\frac{r^2}{S}A^\prime\right)^\prime&=\frac{1}{\mu S}[W_+,[A,W_-]].
	\end{split}
\end{equation}
Integrating and rearranging gives
\begin{equation}\label{intre}
	\begin{split}
		\left(\mu SW_+^\prime\right)\Big|_{r=r_1}&=\left(\mu SW_+^\prime\right)\Big|_{r=r_0}+\int\limits_{r_0}^{r_1}\left(\frac{1}{\mu S}[A,[A,W_+]]-\frac{S\mc{F}}{r^2}\right)dr,\\
		\frac{r_1^2}{S_1}A^\prime|_{r=r_1}&=\frac{r_0^2}{S_0}A^\prime|_{r=r_0}+\int\limits_{r_0}^{r_1}\frac{1}{\mu S}[W_+,[A,W_-]]dr.
	\end{split}
\end{equation}
The right-hand sides of both equations in \eqref{intre} contain only bounded functions, so we can finally conclude that $W_+^\prime$ and $A^\prime$ are bounded on $\bar{\mc{Q}}$. Hence, all field variables are bounded on $\bar{\mc{Q}}$, and if we choose $r_1$ arbitrarily large, we may consider ourselves in the asymptotic regime. $\Box$

\subsection{Global existence of solutions in a neighbourhood of embedded solutions}\label{sec:GloExArg}

One of the major results of this paper is the following theorem. The gist of it is that global solutions to the field equations \eqref{EEs}, \eqref{YMEs}, which we have proven are uniquely characterised by the appropriate boundary values and analytic in those values, exist in open sets of the initial parameter space; and thus that solutions which begin sufficiently close to existing solutions to the field equations will remain close to them as they are integrated out arbitrarily far into the asymptotic regime, remaining regular throughout the range. It can be noted that this argument is extremely similar to those we have crafted for $\sun$ cases \cite{baxter_existence_2008, baxter_existence_2016}, but we give the full proof anyway.

\begin{thr}\label{prop:gloex}
	Let us fix $r_h$ (for black holes only) and $\ell$, and define $r_0=r_h$ for black holes and $r_0=0$ for solitons. Assume we have an existing solution of the field equations \eqref{EEs} and \eqref{YMEs}, with each gauge field function $\omega_j(r)$ possessing $n_j$ nodes each, and with gauge field values $\mc{R}\equiv\{\mc{E}^\prime_{j,h},\omega_{j,h}\}$ for black holes or $\mc{R}\equiv\{\tilde{\psi}_{j},\tilde{u}_{j}\}$ for solitons, with $j=1,..,\mc{L}$, at $r=r_0$. Then all initial gauge field values in a neighbourhood $\mc{R}$ will also give a solution to the field equations in which all the gauge field function $\omega_j(r)$ also has $n_j$ nodes.
\end{thr}
\vspace{5mm}
\textbf{Proof} Assume we have an existing solution to the field equations \eqref{EEs} and \eqref{YMEs}, where each $\omega_j(r)$ has $n_j$ nodes. This solution will have $\mu(r)>0$ for all $r\in[r_0,\infty)$.  From its set $\mc{R}$ of initial conditions, Proposition \ref{prop:gloreg} shows that as long as $\mu(r)>0$ we may integrate this solution out arbitrarily far into the asymptotic regime, which according to and Section \ref{sec:asym} will remain regular and satisfy the boundary conditions as $r\rar\infty$.

From the local existence results (Propositions \ref{prop:lex0}, \ref{prop:lexrh}), we know that for any set of initial values $\mc{R}$ there is a solution locally near $r=r_0$, and that all such solutions are analytic in $\mc{R}$. By analyticity, all sufficiently nearby solutions will have $\mu(r)>0$ for all $r\in\bar{\mc{Q}}\equiv[r_0,r_1]$ for some $r=r_1$ with $r_0<r_1<\infty$. By Proposition \ref{prop:gloreg}, this nearby solution will also be regular on $\bar{\mc{Q}}$.

Now, we let $r_1 >> r_0$ so we are in the asymptotic regime. If $r_1$ is large enough, then $m(r_1)/r_1<<1$ for the existing solution. Let $\tilde{\mc{R}}$  be a different set of initial conditions in some neighbourhood of $\mc{R}$ for gauge fields $\tilde{\omega}_j$, $\tilde{\mc{E}}_j$; and let $\check{m}(r)$ be the mass function of that solution. By analyticity, $\tilde{\mu}(r)>0$ for all $r\in\bar{\mc{Q}}$, so the nearby solution will also be regular on $\bar{\mc{Q}}$.

Also $\tilde{m}(r_1)/r_1<<1$ again due to analyticity, and since $r_1 >> r_0$ we are in the asymptotic regime. If $r_1$ is large enough (and hence $\tau_1$ is very small), the solution will not move very far along its phase plane trajectory as we take $r_1\rar\infty$. Hence, $\tilde{m}(r)/r$ remains small, the asymptotic regime remains valid. According to Proposition \ref{prop:lexinf} and Section \ref{sec:asym}, the solution will reach one of the existing sets of arbitrary asymptotic boundary conditions. Therefore the solution evolved from $\tilde{\mc{R}}$ will be globally regular, exist locally as $r\rar\infty$, and the gauge functions $\tilde{\omega}_j$ will still each have $n_j$ nodes.$\Box$

\begin{cor}
	Nodeless non-trivial solutions to the field equations, i.e. for which $\omega_j(r)\neq0\,\,\forall r$, exist in some neighbourhood of existing trivial SadS solutions, and embedded $\essu$ solutions \eqref{su2embed}, described in Section \ref{sec:SadS}. We emphasise that the functions $\mc{E}_j$ are guaranteed by Section \ref{sec:MonE} to have only one zero, at $r=r_0$.
\end{cor}

\subsection{Existence of solutions in the large $|\Lambda|$ limit ($\ell\rar 0$)}\label{sec:l0}

We have concentrated on finding nodeless solutions largely because in the case of $\sun$, it is known \cite{baxter_stability_2015, nolan_stability_2016} that nodelessness is necessary (but not sufficient) for stability. However it is also seen that another necessary condition was that the absolute value of the cosmological constant  $|\Lambda|\rar\infty$, so that the gravitational sector was stable. Also, numerical results show that as $N$ gets larger, the initial value space for regular solutions shrinks, but for $|\Lambda|\rar\infty$, all solutions are nodeless \cite{baxter_abundant_2008, baxter_existence_2008,baxter_soliton_2007}. Therefore, we finish this work by proving that nodeless black hole and soliton solutions can be found in the limit $|\Lambda|\rar\infty$, i.e. as $\ell\rar0$. 

The strategy is this: We transform to new field variables to find a unique solution to the equations for $\ell=0$, being careful to take this limit correctly. We note that we only need transform the results of Proposition \ref{prop:lexinf} into our new variables, and show that the arguments used in Section \ref{sec:GloExArg} may be easily adapted to serve in a neighbourhood of $\ell\rar 0$. We finally emphasise that we cannot prove the existence of global non-trivial solutions for $\ell=0$, since in that case the asymptotic variable we used in Section \ref{sec:asym} (and here) becomes singular.
\begin{thr}\label{thr:l0}
	For fixed $r_h$, there exist non-trivial solutions to the field equations \eqref{EEs}, \eqref{YMEs}, analytic in some neighbourhood of $\ell=0$, for any choice of boundary gauge field values given by $\{W_+(r_h),\overline{E}_+'(r_h)\}$ (or alternatively, in the base \eqref{rhbase}, $\{\omega_{j,h},\mc{E}_{j,h}\}$, $j=1,...,\mc{L}$) for a black hole, or for a soliton, $\{W_+(0), \overline{E}_+'(0)\}$ (i.e. $\{\tilde{u}_j,\tilde{\psi}_j\}$, $j=1,...,\mc{L}$).
\end{thr}
\textbf{Proof} In the purely magnetic case, we let $\bar{m}=m\ell^2$ and $W_{\pm}^\prime=\frac{\ell}{\sqrt{2}}X_{\pm}$. In that case, we found the unique solution to the field equations $m(r)=r_h^3/2$, $\omega_j(r)=\omega_{j,h}$, $S=1$. In this case, we may note that the purely magnetic solution will satisfy the dyonic system if and only if we also have $\mc{E}_\alpha=0$ ($\forall\alpha\in\Sigma$); so we merely append $\mc{E}_\alpha=0$ to the purely magnetic solution and it is clear that we have a (trivially) dyonic solution. Using the usual basis for $W_+(r)$ \eqref{wBasis} and $\overline{E}_+$ \eqref{Ebar}, the solution is therefore
\begin{equation}\label{solnl0}
	m(r)=\frac{r_h^3}{2},\qquad S(r)=1,\qquad \omega_\alpha(r)=\omega_{\alpha,h},\qquad\mc{E}_\alpha(r)=0,
\end{equation}
for all $r$ and for all $\alpha\in\Sigma$. We note that this is identical to the $\sun$ case, and we treat it similarly. We reprise Proposition \eqref{prop:lexinf} with a change of variables:
\begin{equation}
	\tilde{\lambda}\equiv\lambda\ell^2,\qquad\tilde{m}\equiv m\ell^2,\qquad\tilde{\mu}\equiv\mu\ell^2.
\end{equation}
Then equations \eqref{infeqs} are altered to
\begin{equation}
	\begin{split}
		z\frac{dS}{dz}&=2z^4\left(\norm{v_+}^2S+\frac{\ell^4}{\tilde{\mu}^2S^2}\hat{\zeta}(A,W_+)\right),\\
		z\frac{d\tilde{\lambda}}{dz}=&-2z\left(-\frac{\ell^2\norm{v_0}^2}{s^2}+\frac{\ell^4\hat{\zeta}(A,W_+)}{\mu zS^2}+\ell^2\hat{P}(W_+)+\left(k\ell^2z^2-\tilde{\lambda} z^3+1\right)\frac{\norm{v_+}^2}{2}\right),\\
		z\frac{dv_+}{dz}&=2v_+\left(\frac{1}{\tilde{\mu}z}-1\right)+\frac{1}{\tilde{\mu}z}\left(\mc{F}\ell^2+v_+\left(\tilde{\lambda}z^2-2\hat{P}(W_+)\ell^2z^3+\frac{2\ell^2z^3}{S^2}\norm{v_0}^2\right)\right),\\
		z\frac{dv_0}{dz}&=v_0\left(-z^4\norm{v_+}^2-\frac{\ell^4\hat{\zeta}(A,W_+)}{2\tilde{\mu}^2 S^2}\right)-\frac{\ell^2}{\tilde{\mu}z}\hat{\mc{Y}}(A,W_+).
	\end{split}
\end{equation}
The terms involving $\ell$ are $O(z)$ or higher, so that the argument given in Section \ref{sec:LocExInf} carries across unchanged. Thus for arbitrarily small $\ell$, we may find solutions that exist locally in the limit $r\rar\infty$.

The argument that proves that non-trivial black hole solutions exist globally in this regime is very similar to Proposition \ref{prop:gloex}. We fix $r_h$, take the existing trivial solution \eqref{solnl0} (with initial conditions $\{\omega_{j,h}\}$ non-zero in general), and consider varying $\{\omega_{j,h}\}$, and varying both $\ell$ and $\{\mc{E}'_{j,h}\}$ away from zero. Note that for the embedded solution \eqref{solnl0}, all magnetic gauge fields will be nodeless. We choose some $r_1 >> r_h$ so that we can consider $r_1$ in the asymptotic regime. Propositions \ref{prop:lexrh} and \ref{prop:gloreg} imply that for $\ell$ sufficiently small we may find new solutions near the trivial solution, which will begin regularly near $r = r_h$ and remain regular throughout $(r_h, r_1]$. Once we are in the asymptotic regime, we can use Section \ref{sec:asym} to prove that: the solution will remain regular as $r\rar\infty$; that all $\omega_j(r)$ will be nodeless; and given that $\mc{E}_j(r)$ are all positive and monotonic, that all the $\mc{E}_j(r)$ will likewise be nodeless for $r>r_h$. 

The corresponding argument for solitons is similar in form to that for black holes, and as in the purely magnetic case, we must be careful about how we take the limit $\ell\rar0$ due to the unboundedness here of the parameter $\tau=\ell r^{-1}$ that we used for black holes -- hence we take the co-ordinate $x=\ell^{-1}r$. Again we use the unique solution obtained in \cite{baxter_general_2016} and append $\mc{E}_j\equiv 0$, i.e. $\hat{\psi}_j=0$, giving
\begin{equation}\label{solnl0sol}
	m(x)\equiv 0,\quad S(x)\equiv 1,\quad\hat{\psi}_j(x)\equiv 0,\quad\hat{u}_j(x)=\tilde{u}_j\left\{\,_2 F_1 \left(\frac{\kappa_j+1}{2},\frac{\kappa_j}{2};\frac{2\kappa_j+1}{2};-x^2\right)\right\},
\end{equation}
for all $x$ and for all $j\in\{1,...,\mc{L}\}$, and $\{\kappa_j\}$ being the sequence of integers defined in Section \ref{sec:sl2c}, which integers depend on the group $G$.

We proceed in a very similar fashion to the black hole case. We fix $r_h$, take the existing solution \eqref{solnl0sol} with arbitrary $\tilde{u}_{j}\,\,\forall j$, and consider varying $\{\tilde{u}_{j,}\}$, varying $\ell$ away from 0, and varying $\{\tilde{\psi}_j\}$ away from 0. Note again that for the embedded solution \eqref{solnl0}, all magnetic gauge fields will be nodeless. Following identical logic to the black hole case, we choose $r_1 >> 0$, so that Propositions \ref{prop:lex0} and \ref{prop:gloreg} confirm that for $\ell$ and $\{\tilde{\psi}_j\}$ sufficiently small we can find solutions near the existing unique solution which will begin regularly near $r = 0$, remain regular until the asymptotic regime (by Section \ref{sec:asym}), and meet up with regular boundary conditions at infinity. Furthermore, these neighbouring solutions will once again have $\omega_j(r)\neq0\,\,\forall r$, and $\mc{E}_j(r)\neq0\,\,\forall r>0$. $\Box$

It is finally worth noting that in varying $\mc{E}_j$ away from 0, we are requiring that these solutions have a small electric field; the magnetic field, as in the purely magnetic case, is more arbitrary.

\section{Conclusions}\label{sec:conc}

The purpose of this work was to investigate global solutions to 4D static adS EYM equations, for topologically symmetric black holes and solitons, with non-trivial electric and magnetic sectors. We began by deriving the correct form for the connection and metric in our case (Section \ref{sec:ansatze}). Then, we used these to derive the field equations in a very general case, before reducing them down to the regular case (Section \ref{sec:FEs}). We saw that the equations reduced once more to looking similar to the $\mk{su}(N)$ equations studied in \cite{baxter_existence_2016}. In Section \ref{sec:embed}, we found some trivial embedded solutions corresponding to previously proven solutions from \cite{winstanley_existence_1999}. After that in Section \ref{sec:LocEx}, for each of the boundaries $r=0$, $r=r_h$ and $r\rar\infty$ in turn, we analysed the boundary conditions, and then used a well-known theorem of differential equations (Theorem \ref{Kthm3}) to establish the existence of solutions close to each boundary which are analytic in their boundary values.

We proceeded by `stitching' the solutions together, using a series of proofs which showed that if we begin a solution with some arbitrary boundary values at the event horizon (or the origin for solitons), then we may continue to integrate the solution into the asymptotic regime, and it will remain regular (Section \ref{sec:GloRegArg}). Proposition \ref{prop:lexinf} implied that solutions will exist near infinity that the evolved solution will match up to, confirming results from \ref{sec:LocEx} concerning the lack of constraints on the boundary conditions here. To finish the proof of global solutions, we argued in Section \ref{sec:GloExArg} that given the analyticity of the boundary values, any solution which starts sufficiently nearby a trivial/embedded solution will stay nearby it into the asymptotic regime, where it will remain regular and represent a new non-trivial solution. Importantly for later work, this nearby solution will have all $\omega_j(r)$ nodeless for all $r$, and all $\mc{E}_j(r)$ nodeless for all $r>r_h$ (or $r>0$ for solitons). Finally, using appropriate variable changes, we gave a similar argument which establishes the existence of nodeless solutions in the regime $|\Lambda|\rar\infty$ (Section \ref{sec:l0}).

The main result in this paper is the proof of the existence of non-trivial global nodeless solutions to 4D dyonic adS EYM theories for semi-simple, compact and simply-connected Lie gauge groups, both in neighbourhoods of existing (embedded) solutions and in the regime where $|\Lambda|\rar\infty$. This is not an unexpected result, but it is a nice one, since the author believes this represents the most general model of 4D static adS EYM systematically studied to date. The fact that this regular case bears similarity to the $\mk{su}(N)$ model is handy; the main difference between the regular $\sun$ and general cases is the Cartan matrix used. This means that while numerical results for other Lie groups will in general be different to those discovered for $\sun$ \cite{baxter_abundant_2008,baxter_topological_2016,baxter_existence_2008,baxter_soliton_2007}, they should bear some structural similarities to the $\sun$ case -- certainly we expect solutions to be found in continuous bands in the initial value space (possibly given some bounds on the values of $\Lambda$ and $r_h$), also giving hope that some of these solutions will be stable since perturbed solutions may be able to find nearby regular boundary values. This may form the subject of a future investigation.

There are a few more natural extensions this work suggests. Bizon's ``no-hair'' theorem \cite{bizon_colored_1990} states that, ``In any given matter model, a \emph{stable} black hole is characterised by a finite number of unique charges". In addition, the solutions we found were nodeless, which in the purely magnetic $\sun$ case \cite{baxter_stability_2015} and the dyonic $\essu$ case \cite{nolan_stability_2016}, was a necessary requirement for stability, as was $|\Lambda|\rar\infty$. In light of the `No-hair' theorem and these facts, this suggest a stability analysis is necessary -- however, in $\essu$ the lack of an obvious simplifying global gauge meant the system was extremely intricate \cite{nolan_stability_2016} and proving stability will be exceptionally difficult, so an investigation of the linear stability of the $\sun$ dyonic system might be a better, if almost equally difficult, first step. 

Also, it is known that topological dyonic $\sun$ models are a good model for holographic semiconductors via the adS/CFT correspondence, which states that gravitational results in adS can be translated into QFT results on the (Minkowski) boundary. It is interesting to wonder whether this larger class of models also has applications to condensed matter physics. Finally, it would be of great interest to know if this work could be used at all in the investigation of the Black Hole Information Paradox, in light of some of Hawking's recent comments \cite{hawking_soft_2016} on how black hole hair may be used to resolve the problem of information loss in black hole spacetimes. Such questions provide ample possibilities for future research.

\vspace{1cm}
\hrule
\appendix

\section*{References}
%
%
%
\bibliographystyle{unsrt}

\end{document}